\documentclass[12pt,preprint]{aastex}
\received{}
\revised{}
\accepted{}
\usepackage{amsmath}

\journalid{}{}
\articleid{}{}
\paperid{}
\shortauthors{Perrett et al.}
\shorttitle{SNLS RTA and Malmquist Bias}

\def\Aunit{\mathrel{\hbox{$\times\,10^{-14}\,\mathrm{SNe\,yr}^{-1}\,M_\odot^{-1}$}}}
\def\Bunit{\mathrel{\hbox{$\times\,10^{-4}\,\mathrm{SNe\,yr}^{-1}\,(M_\odot\,\mathrm{yr}^{-1})^{-1}$}}}

\begin{document}

\title{Real-time Analysis and Selection Biases in the Supernova Legacy
  Survey\footnotemark[1]}

\author{
K.~Perrett\altaffilmark{2,3},
D.~Balam\altaffilmark{4},
M.~Sullivan\altaffilmark{5},
C.~Pritchet\altaffilmark{6}, 
A.~Conley\altaffilmark{2,7},
R.~Carlberg\altaffilmark{2},
P.~Astier\altaffilmark{8},
C.~Balland\altaffilmark{8},
S.~Basa\altaffilmark{9},
D.~Fouchez\altaffilmark{10},
J.~Guy\altaffilmark{8},
D.~Hardin\altaffilmark{8},
I.~M.~Hook\altaffilmark{4,11},
D.~A.~Howell\altaffilmark{12,13},
R.~Pain\altaffilmark{8}, 
N.~Regnault\altaffilmark{8}
}
\email{perrett@astro.utoronto.ca, sullivan@astro.ox.ac.uk}
\footnotetext[1]{
 Based on observations obtained with MegaPrime/MegaCam, a joint project
 of CFHT and CEA/DAPNIA, at the Canada-France-Hawaii Telescope (CFHT)
 which is operated by the National Research Council (NRC) of Canada,
 the Institut National des Sciences de l'Univers of the Centre National
 de la Recherche Scientifique (CNRS) of France, and the University of
 Hawaii. This work is based in part on data products produced at the
 Canadian Astronomy Data Centre as part of the Canada-France-Hawaii
 Telescope Legacy Survey, a collaborative project of NRC and CNRS.
}

\altaffiltext{2}{Department of Astronomy and Astrophysics, University
  of Toronto, 50 St. George Street, Toronto, ON, M5S 3H4, Canada}

\altaffiltext{3}{Network Information Operations, DRDC Ottawa, 3701
  Carling Avenue, Ottawa, ON, K1A~0Z4, Canada}

\altaffiltext{4}{Dominion Astrophysical Observatory, Herzberg
  Institute of Astrophysics, 5071 West Saanich Road, Victoria, BC,
  V9E~2E7, Canada}

\altaffiltext{5}{Department of Physics (Astrophysics), University of
  Oxford, DWB, Keble Road, Oxford OX1 3RH, UK}

\altaffiltext{6}{Department of Physics \& Astronomy, University of
  Victoria, PO Box 3055, Stn CSC, Victoria, BC, V8W~3P6, Canada}

\altaffiltext{7}{Center for Astrophysics and Space Astronomy, 
  University of Colorado, 593 UCB, Boulder, CO, 80309-0593, USA}

\altaffiltext{8}{LPNHE, Universit\'{e} Pierre et Marie Curie Paris 6, Universit\'{e} Paris Diderot Paris 7, CNRS-IN2P3, 4 place Jussieu, 75005 Paris}

\altaffiltext{9}{Laboratoire d'Astrophysique de Marseille, P\^{o}le de
  l'Etoile Site de Ch\^{a}teau-Gombert, 38, rue Fr\'{e}d\'{e}ric
  Joliot-Curie, 13388 Marseille cedex 13, France}

\altaffiltext{10}{CPPM, CNRS-IN2P3 and University Aix Marseille II,
  Case 907, 13288 Marseille cedex 9, France}

\altaffiltext{11}{INAF, Osservatorio Astronomico di Roma, via Frascati
  33, 00040 Monteporzio (RM), Italy}

\altaffiltext{12}{Las Cumbres Observatory Global Telescope Network,
  6740 Cortona Dr., Suite 102, Goleta, CA 93117, USA}

\altaffiltext{13}{Department of Physics, University of California,
  Santa Barbara, Broida Hall, Mail Code 9530, Santa Barbara, CA
  93106-9530, USA}

\begin{abstract}
  The Supernova Legacy Survey (SNLS) has produced a high-quality,
  homogeneous sample of Type~Ia supernovae (SNe~Ia) out to redshifts
  greater than $z=1$.  In its first four years of full operation (to
  June 2007), the SNLS discovered more than $3000$ transient
  candidates, $373$ of which have been confirmed spectroscopically as
  SNe~Ia. Use of these SNe Ia in precision cosmology critically
  depends on an analysis of the observational biases incurred in the
  SNLS survey due to the incomplete sampling of the underlying SN~Ia
  population.  This paper describes our real-time supernova detection
  and analysis procedures, and uses detailed Monte Carlo simulations
  to examine the effects of Malmquist bias and spectroscopic sampling.
  Such sampling effects are found to become apparent at $z\sim 0.6$,
  with a significant shift in the average magnitude of the
  spectroscopically confirmed SN~Ia sample towards brighter values for
  $z\ga 0.75$.  We describe our approach to correct for these
  selection biases in our three-year SNLS cosmological analysis
  (SNLS3), and present a breakdown of the systematic uncertainties
  involved.
\end{abstract}


\keywords{cosmology: observations --- supernovae: general --- methods:
  data analysis --- techniques: photometric --- surveys}

\section{Introduction}

Type Ia supernovae (SNe~Ia) have frequently demonstrated their value
as standardizable candles out to redshifts of $z\ga 1.5$. Studies of
the expansion rate of the universe using SNe~Ia have revealed the
cosmic acceleration \citep{per99,rie98}, providing the most direct
evidence for the presence of Dark Energy needed to drive the expansion
of the universe.  Subsequent searches have led to SN~Ia discoveries
over a wide range of redshifts
\citep[e.g.,][]{ast06,mik07,sak08,rie07}.

SNe~Ia are relatively rare, transient objects lasting a few weeks in
the rest-frame. They are most efficiently found through photometric
monitoring of large areas of sky using a ``rolling-search'' technique
to both find and follow SN events. Current examples include the
ESSENCE survey \citep{mik07,wv07}, the SDSS-II SN Survey
\citep{sak08,kes09}, and the Supernova Legacy Survey
\citep[SNLS;][]{ast06}. The SNLS observed four 1-square-degree Deep
Fields as part of Canada-France-Hawaii Telescope Legacy Survey
(CFHT-LS)\footnote{http://www.cfht.hawaii.edu/Science/CFHLS/}.
Multi-color light curves were obtained using the MegaCam imager on
CFHT, with follow-up spectroscopy from the Gemini, VLT, and Keck
telescopes.  Owing in large part to its careful survey design and a
substantial commitment of observing time from CFHT and the follow-up
telescopes, the SNLS now offers a large, consistent, and well-defined
set of SNe~Ia measured out to $z\sim 1$.

Real-time supernova searches require adequate light-curve sampling, a
rapid output of photometry measurements, and a method to quickly
prioritize candidate lists for follow-up studies
\citep[e.g.,][]{sul06a,mik07,sak08}. It is most advantageous to obtain
spectroscopy near maximum SN brightness \citep[e.g., see][]{how05}.
Since SNe~Ia have typical rise times of $\sim 19$ days
\citep[rest-frame $B$, ][]{con06}, and considering that observations
are typically only taken during dark and grey time, in some cases only
a couple of measurable epochs are available before peak.  Data
organization and the coordination of follow-up activities are
therefore critical.  In the SNLS, Canadian and French teams each ran
independent detection pipelines, then amalgamated both candidate lists
into a centralized database. Herein, we report exclusively on the
Canadian real-time analysis pipeline.

As the extent and quality of SN~Ia data sets improve, the relative
impact of statistical uncertainties decreases and the management of
systematic errors becomes the limiting factor
\citep{kes09,reg09,con10}.  One important source of systematic error
in SN~Ia studies arises from the incomplete sampling of the underlying
SN~Ia population: surveys tend to preferentially discover and follow
brighter objects at higher redshifts, leading to Malmquist bias
\citep{mal36} and other selection effects.  Using detailed Monte Carlo
simulations of artificial SNe~Ia in all of the real-time detection
images, we can determine the recovery efficiency and calculate the
magnitude of the various sampling biases \citep[see also][]{con10}.
These can then either be directly applied to the data, or used to
motivate fitting priors or corrections used when determining
cosmological parameters \citep[e.g.,][]{kes09}.

This paper is organized as follows: An overview of the SNLS and a
description of the Canadian SNLS real-time-analysis (RTA) pipeline and
web database are given in \S\ref{sec:RTA}. \S\ref{sec:followup}
presents the methods used in ranking candidates for spectroscopic
follow-up observations.  The Monte Carlo simulations used to calculate
the SNLS detection efficiencies are described in \S\ref{sec:comp}
along with the recovery results.  The effects of Malmquist bias and
spectroscopic sampling are explored in \S\ref{sec:malm}, with a
discussion of the systematic errors in the Malmquist bias in
\S\ref{sec:sys}.  The paper concludes with a summary in
\S\ref{sec:summary}.

\section{Observations and real-time analysis}
\label{sec:RTA}

\subsection{SNLS observations}
\label{sec:obs}

\begin{deluxetable}{cccl}
\tablewidth{0pt}
\tablecaption{Deep Field centers\label{tab:fields}}
\tablehead{
  \colhead{Field} & \colhead{RA (J2000)} & \colhead{DEC (J2000)} & 
  \colhead{Overlapping Surveys}
}
\startdata
D1 & 02:26:00.00 & -04:30:00.0 & VIMOS, SWIRE, GALEX, XMM Deep\\
D2 & 10:00:28.60 &  02:12:21.0 & COSMOS/ACS, VIMOS, SIRTF, GALEX, XMM\\
D3 & 14:19:28.01 &  52:40:41.0 & (Groth Strip) ACS, DEEP-II, GALEX, AEGIS\\
D4 & 22:15:31.67 & -17:44:05.7 & XMM Deep\enddata
\end{deluxetable}

The five-year Supernova Legacy Survey ran from 2003-2008 and was based
on images obtained from the CFHT-LS Deep Synoptic Survey.  The four
Deep Field pointings were positioned to avoid extremely bright stars,
minimize Milky Way extinction, and overlap with the fields of other
multi-wavelength surveys to provide complementary data.  The
coordinates of the SNLS Deep Field centers are given in
Table~\ref{tab:fields}.

Observations were obtained in queue-scheduled mode at CFHT using the
MegaCam wide-field imager \citep{bou03}.  MegaCam consists of 36
$2048\times4612$ pixel$^2$ CCDs arranged in a $4\times9$ square
mosaic.  The CCDs have a resolution of $\sim0.186\arcsec$/pix and
provide good sampling of the $0.7\arcsec$ median seeing at CFHT.  A
large ($1.5\arcmin$) dithering pattern was used to cover the
$80\arcsec$ wide gaps between CCD rows in the MegaCam mosaic.

The SNLS employed a rolling-search observing strategy, with target
fields imaged every $3-4$ days in dark or grey time using multiple
filters to provide optimal light-curve sampling and the opportunity
for early SN discovery.  This cadence translates to $\sim2-3$ days in
the SN rest frame.  Each MegaCam queue run lasted on average $16-17$
nights centered around the new moon, typically providing 5 epochs
(nights) of observation per field per lunation.  An advantage of this
scheme is that the same telescope is used to discover and follow-up
all candidates, leading to considerable advantages in calibration.

\begin{deluxetable}{lrl}
\tablewidth{0pt}
\tablecaption{Typical SNLS/Deep exposures\label{tab:exptimes}}
\tablehead{\colhead{Filter} & \colhead{ExpTime (sec)} & \colhead{Epoch Number}}
\startdata
$g_M$         & $5 \times 225 = 1125$ & 1,2,3,4,5\tablenotemark{a}\\ 
$r_M$         & $5 \times 300 = 1500$  & 1,2,3,4,5\\ 
$i_M$ (long)  & $7 \times 520 = 3640$  & 1,3,5    \\ 
$i_M$ (short) & $5 \times 360 = 1800$  & 2,4      \\ 
$z_M$         & $10 \times 360 = 3600$  & 1,3,5      
\enddata
\tablenotetext{a}{$g_M$ acquisition depended on Moon brightness.}
\end{deluxetable}

The four filters used by SNLS --- $g_Mr_Mi_Mz_M$ --- are similar (but
not identical) to the filter system adopted by the Sloan Digital Sky
Survey \citep{fuk96,smi02}.  Deep Field images were also obtained in
$u_M$, but these were not time-sequenced and thus were not used in the
real-time analysis.  A detailed description of the SNLS photometric
calibration is provided in \citet{reg09}.  Observations in the primary
search filter ($i_M$) were alternated between ``long'' and ``short''
epochs in order to balance the requirements for adequate light-curve
coverage with the need to conduct several deep searches within each
queue run.  Filter spacing and nightly exposure times for a typical
queue run are provided in Table~\ref{tab:exptimes}.

A major strength of the SNLS program is that it obtained better filter
coverage and light-curve cadence than had been achieved by previous
high-$z$ surveys.  Other advantages included having a large quantity
of dedicated telescope time with data acquisition coordinated through
CFHT's Queued Service Observing mode.  Furthermore, developing a
careful strategy for the organization, analysis, and communication of
SNLS data products was necessary to efficiently manage the continual
acquisition of new target information.  The process employed by the
SNLS collaboration included linking remote search pipelines with a
centralized database, and providing web-based applications as a simple
user interface with communication and analysis functionalities.  These
are described in the next section.

\subsection{Image processing}
\label{sec:pipeline}

MegaCam images of the Deep Fields destined for SNLS analysis were
pre-processed in real-time using the CFHT Elixir pipeline
\citep{mag04}.  The Elixir pipeline performed bias subtraction and
flat-fielding, as well as a basic fringe subtraction on the $i_M$ and
$z_M$ images.  Elixir-processed images were promptly made available to
the SNLS team for further processing.

\begin{figure}
\plotone{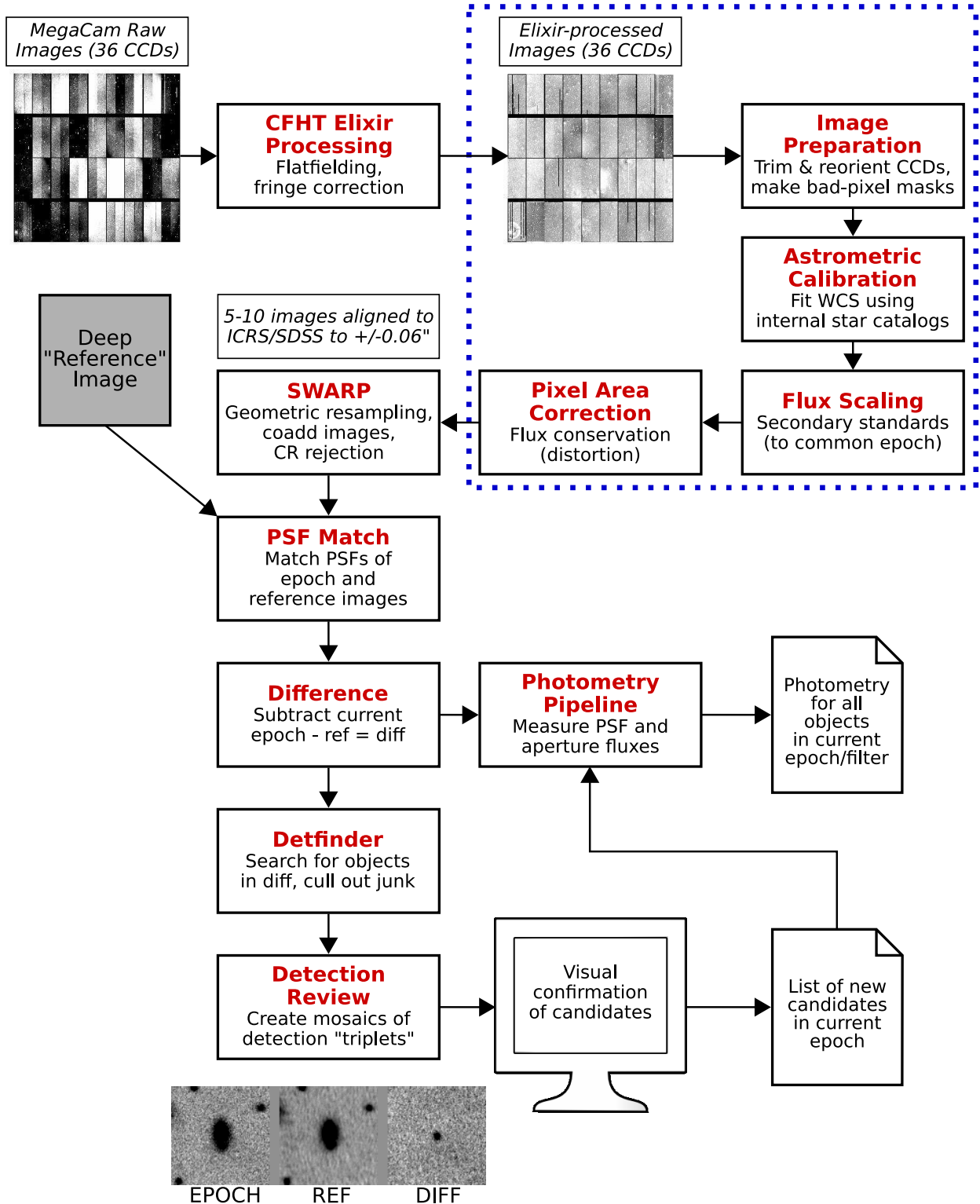}
\caption{The Canadian RTA pipeline.  The dotted box delineates the
  procedures that monitor incoming Elixir-processed images.  The rest
  of the steps proceed on the pre-processed images as described in the
  text.}
\label{fig:pipeline}
\end{figure}

The real-time detection pipeline implemented by the Canadian SNLS team
is outlined in the flowchart of Figure~\ref{fig:pipeline}.  Many of
the processing steps were accomplished using IRAF\footnote{Image
  Reduction and Analysis Facility (IRAF) is distributed by the
  National Optical Astronomy Observatory, which is operated by the
  Association of Universities for Research in Astronomy (AURA) under
  cooperative agreement with the National Science Foundation.} tasks.
Due to time constraints and observatory requirements, the image
processing was carried out remotely on SNLS computers at the CFHT
headquarters in Waimea.  Certain shortcuts were taken in the real-time
pipeline as compared with the ``final'' photometry procedure used in
the cosmological analysis \citep[described in][]{guy10}.  These were
necessary since the SNLS RTA was operating under significant time
pressures for discoveries and analysis.

Elixir-processed images were trimmed and reoriented, then bad pixel
masks were created to flag and remove image artifacts and saturated
stars.  To account for optical distortions, astrometric solutions were
fit to each CCD using combined catalogs from USNO-B and SDSS to align
the images to the International Celestial Reference System (ICRS). The
resulting fits had RMS residuals that were typically better than $\la
0.06\arcsec$ (0.3 pixels) across the mosaic.

Photometric alignment of the RTA images was accomplished using
aperture flux measurements of $\sim 20\,000 - 40\,000$ ``RTA secondary
standards'' selected over a wide range in brightness for all filters
within each field.  These secondaries were chosen from isolated
objects exhibiting no significant changes in brightness over many
epochs.  Comparing the measured fluxes for the secondaries in a given
epoch to their catalog values provided a multiplicative scaling factor
used to flux calibrate each image. As such, all flux measurements over
the duration of any SN light curve were consistently in the same
reference system, removing variations in exposure time, airmass, and
extinction. In the RTA, the images were calibrated to the AB
photometric system \citep{oke83}; this differs from the method used in
final photometry, which calibrated to the \citet{lan92} photometric
system \citep{reg09}.

Individual flux-scaled CCD images were adjusted for variations in
effective pixel area using the IRAF MSCRED (mosaic data reduction)
task MSCPIXAREA.  Resampling and co-addition of the night's images
were then done using the TERAPIX SWarp\footnote{\tt
  http://astromatic.net} program to form a median mosaic image and its
accompanying weight-map for each field$+$filter combination.

The subtraction of a deep reference image from the science mosaic
revealed the presence of SNe and other variable objects in the data.
References for each field were constructed by selecting photometric
images with excellent seeing ($\la 0.5\arcsec$) and good transparency.
To avoid the presence of SN light in the reference images, only frames
taken in years prior to the current observations were included in each
stack\footnote{RTA detections made during the first year targeting
  each field were made using whatever prior data was available.  The
  quality of the reference stacks improved with each successive
  observing season.}.  These component images were flux scaled,
WCS-aligned and co-added as described above to form the reference
image.  During the course of the survey, new RTA stacks were
constructed at least once per season as more images became available
to improve the depth of the reference mosaic.

Differences in image quality between the science and reference mosaics
must be taken into account prior to image subtraction.  A
non-parametric point-spread function (PSF) matching routine was used
to effectively degrade the seeing of the better quality image ---
typically the reference --- to match the seeing of the other image
\citep[see][Appendix A]{fra08}.  A Moffat function was fit to a select
set of a few thousand isolated, bright ``PSF stars'', culling those
with poor-quality fits.  A variable convolution kernel was computed
for each star, matching the two PSFs.  The reference image was then
convolved with the kernel to produce a PSF-matched mosaic.  Finally,
the matched frame was subtracted from the science frame.

\begin{figure}
\plotone{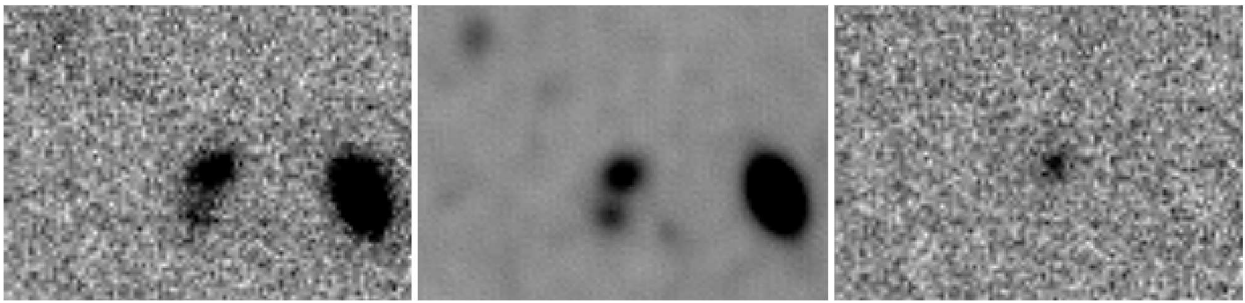}
\caption{A sample detection triplet showing an SN~Ia candidate
  (SNLS-04D2ca) at $z=0.835$.  The science image in the $i_M$ filter
  for the night of 2004-03-10 is shown on the left, the PSF-matched
  $i_M$ reference image is in the center, and the difference between
  the two is the image on the right.  The subtracted mosaic is used
  for candidate detection as described in the text.  Each image is
  19\arcsec wide.}
\label{fig:triplet}
\end{figure}

A sample ``detection triplet'' is presented in
Figure~\ref{fig:triplet} for a high-redshift SN candidate: SNLS-04D2ca
at $z=0.835$.  It shows thumbnail images of corresponding regions in
the science image for one epoch, the PSF-matched reference image, and
the difference.

One drawback of the real-time analysis procedure is that, in the
interest of processing speed, the method of co-adding the individual
epoch images prior to the PSF-matching step meant that if the image
quality changed significantly during a night's set of observations,
the resulting difference image may have contained residuals from poor
subtractions of host galaxies.  Normally, such residuals were easy to
identify and immediately cull from the output detection list, as
described in the next section.

\subsection{Supernova detection}
\label{sec:dets}

Once the host galaxy light was removed via the subtraction of the
reference image, potential supernova candidates appeared in the
subtracted images as star-like objects with positive flux (see the
right-hand panel of Figure~\ref{fig:triplet}).  The subtracted images
were searched using an algorithm that employed the TERAPIX SExtractor
routine \citep{ber96} to build a catalog of possible objects and apply
a series of tunable culling procedures.  Each preliminary catalog of
detections was cleaned of objects with very low signal-to-noise or a
significant fraction of bad pixels, as well as objects with irregular
profiles or high second-order moments.  Adjustable image quality
limits were available for additional culling on PSF-fit quality
parameters and background sky measurements.

Residual features in the difference images could result from poorly
subtracted galaxies or bright stars.  These residuals commonly
appeared as alternating bright and dark regions --- the bright areas
could sometimes survive the other culling procedures.  To compensate
for this, the detection code also compared the image statistics in
defined regions immediately surrounding the candidate to help remove
such false detections from the object list.

The resulting raw list of variable object candidates in each epoch was
then used to produce a composite mosaic of detection triplets for
visual review and a manual selection of candidates.  The number of raw
output detections depended on the parameter limits established for the
cuts (e.g., the S/N cutoff).  We generally opted to use liberal cuts
and do a manual pre-selection of SN candidates from the resulting $\la
200$ raw detections. Of these, typically $\sim 10\%$ remained as
plausible SN candidates. Using the visual display of candidates, many
of the remaining PSF-matching residuals, image artifacts, moving
objects, and stellar cores that may have survived the previous culling
steps could be removed.

Detections were primarily made using the $i_M$ (and to a lesser extent
$r_M$) images in order to focus on the target redshift range of
interest.  Candidates that were too faint in $i_M$ were poorly-suited
for SNLS follow-up spectroscopy and cosmological analysis.  A detailed
analysis of the detection incompleteness is presented later in
\S\ref{sec:comp}.

\subsection{Photometry}
\label{sec:phot}

Supernova flux measurements were calculated using a PSF-fitting
routine.  The same sets of isolated stars used in PSF-matching step
(\S\ref{sec:pipeline}) were used to create a mean local PSF and
residuals map for input to the fitting routine.  The stellar model was
then scaled to match each detection in the difference image, thus
providing the flux of each source.  In addition to the PSF photometry,
aperture fluxes were also calculated using a range of static and
seeing-dependent aperture sizes.

The output PSF-fit fluxes measured in each epoch were used to
construct multi-color light curves for all of the variable-object
candidates detected by SNLS. These light curves were then fit with
SN~Ia templates to assist in classification and to provide photometric
redshift measurements; this process is described in
\S\ref{sec:followup}.

\begin{figure}
\plotone{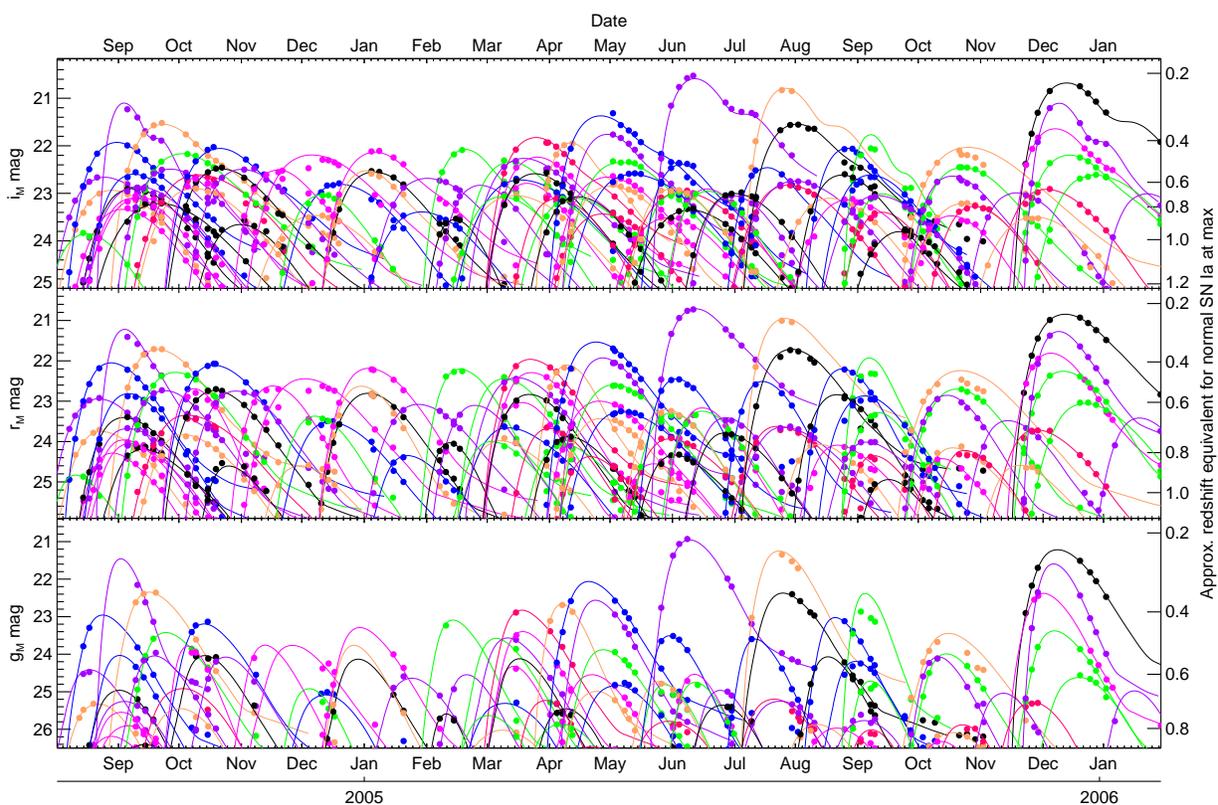}
\caption{Observed light curves for some of the SNLS SN~Ia candidates,
  showing $i_M$ (top), $r_M$ (middle), and $g_M$ (bottom).  Each
  observation is given by a dot on the corresponding SN~Ia light-curve
  fits (solid lines).  The observations were clustered during dark
  time, with additional gaps due to runs of poor weather.}
\label{fig:rollingLC}
\end{figure}

A sample set of SN~Ia light curves in $g_Mr_Mi_M$ is shown in
Figure~\ref{fig:rollingLC}, demonstrating the typical observing
cadence obtained by the SNLS.

\subsection{The SNLS web database}
\label{sec:db}

Once the manual review and culling was complete for a given epoch, the
preliminary detection list was loaded into the SNLS database in
preparation for ranking and follow-up measurements.  New SN candidates
and photometry were made available to the community via the World Wide
Web\footnote{\tt http://legacy.astro.utoronto.ca}, typically within
$4-6$~hours per field.

An observing request utility was used communicate the latest SNLS
requirements to the CFHT queued service observers (QSOs).  This tool
kept the QSOs advised of the detailed priorities of the survey,
reaching a level beyond what could normally be expressed in the Phase
II application that was submitted at the start of each run.  Often,
observing conditions, SN candidate status, and follow-up telescope
availability called for some flexibility in field/filter observing
priorities for SNLS.

Additional online tools facilitated the coordination of observing
schedules and follow-up allocations so as to optimize the use of the
available 8-10~m telescope time.  A detection ``forum'' was used to
collect comments and personalized rankings for candidates being
considered for follow-up spectroscopy.  Automated (but customizable)
online finder charts simplified subsequent observations in both
single-target and multi-object modes.

It is worth noting that the light curves of {\em all} candidates in
the database were followed until they became too faint or were
unequivocally rejected as a non-SNe.  No candidates were thrown away,
no matter how poor; all have been stored in the database for later
studies.

\section{Prioritization and spectroscopy}
\label{sec:followup}

As with most supernova surveys, the SNLS was limited by the amount of
time available for spectroscopic follow-up.  Even with its maximum
allocation of $\sim140$ hours of 8-10~m telescope time per semester,
the SNLS was only able to follow a small fraction of all possible SN
detections.  Differentiating the SNe~Ia from other variable sources
using only a few early observations presented a significant challenge.

\subsection{Candidate ranking}
\label{sec:ranking}

The success of any supernova survey relies on the careful selection of
SN~Ia candidates with enough time to optimally schedule spectroscopic
follow-up close to peak magnitude.  Light-curve fits and brightness
cuts were the primary methods used to identify and rank SNLS targets
for spectroscopy.  The addition of photometric pre-selection
dramatically improved the SNLS spectroscopic efficiency over
comparable surveys, even at high redshift \citep{how05}.

Probable SNe~Ia could typically be identified from two or more epochs
of real-time photometry with the supernova light-curve fitting
technique described in detail in \citet{sul06a}.  The output
light-curve fits are parameterized by redshift, stretch, peak
brightness, and time of maximum light. Comparisons between photometric
and spectroscopic redshifts for the SNLS SNe~Ia are available in
\citet{sul06a} and \citet{per10}.  Non-SNe~Ia could be screened out
based on their measured colors ($g_M-r_M$) and rise times.  All SN
candidates were measured against several criteria to determine their
suitability and prioritization for spectroscopic follow-up.
Candidates with poor-quality SN~Ia light-curve fits, or with fitted
stretch values outside the range $0.7\leq s \leq 1.3$, were generally
given the lowest ranking or were rejected entirely.

Observing near maximum light yields the brightest contrast between the
SN and its host, and is the time when the spectral lines most clearly
differ between supernova types.  SNLS SNe~Ia were normally discovered
before peak, and the predicted phase from the fits allowed for optimal
spectroscopic scheduling.  Each candidate SN was required to have been
detected at most seven days (rest frame) past its estimated peak date
in order to qualify for timely spectroscopic observing.  Objects
discovered at the very start or end of an observing season were
typically discarded if they would have no measurements either before
or after peak. Enough of the candidate's light curve must be measured
to provide adequate fits and prove useful for cosmology.

Limits were placed on the minimum observed brightness and fractional
increase of each candidate compared with its host galaxy.  Percent
increase is defined as:
\begin{equation}
\%\mathrm{inc} = \frac{f_\mathrm{sci} -
  f_\mathrm{ref}}{f_\mathrm{ref}} \times 100\%,
\label{eq:pcinc}
\end{equation}
where flux is measured in small (3-pixel radius) apertures in the
science image ($f_\mathrm{sci}$) and the reference ($f_\mathrm{ref}$).

Due to the vagaries of weather and telescope scheduling, it is
impossible to give a precise description of the selection process for
sending candidates for spectroscopic follow-up.  However, a good
description of the average algorithm is as follows, with all observed
magnitudes presented in the AB system:
\begin{enumerate}
\item Extremely bright SNe ($i_M<22.9$) always qualified for
  spectroscopic follow-up.
\item Moderately bright SNe ($22.9\leq i_M<23.8$) qualified if they
  had \%inc$>30$ in some epoch.
\item Fainter SNe ($23.8\leq i_M<24.4$) must have had \%inc$>100$ in
  some epoch to qualify.
\end{enumerate}
Candidates fainter than $i_M\sim24.4$ were generally never observed
spectroscopically.  

The basic form of these limits was suggested by the experience of
those selecting candidates for follow-up, and tuned by comparing the
distributions of peak magnitude, color, and light-curve width of
moderate redshift SNLS SNe ($z < 0.6$), where the survey should be
essentially complete, to those of higher redshift SNe.

In practice, the selection process remained somewhat subjective.
External factors also played a role in the ranking of candidates by
SNLS members --- e.g., the availability of follow-up telescope time,
detailed scheduling limitations, long-term weather conditions, and the
desired spread in redshifts needed for cosmological measurements.

\subsection{Spectroscopy}
\label{sec:spec}

Spectroscopic observations are necessary to verify candidate types and
to obtain accurate redshift measurements for the SN and/or its host
galaxy.  SNLS spectra were obtained from 8-10~m class telescopes
including Gemini\footnote{This work is based in part on observations
  obtained at the Gemini Observatory, which is operated by the
  Association of Universities for Research in Astronomy, Inc., under a
  cooperative agreement with the NSF on behalf of the Gemini
  partnership: the National Science Foundation (United States), the
  Science and Technology Facilities Council (United Kingdom), the
  National Research Council (Canada), CONICYT (Chile), the Australian
  Research Council (Australia), CNPq (Brazil) and CONICET (Argentina).
  Gemini program IDs: GS-2003B-Q-8, GN-2003B-Q-9, GS-2004A-Q-11,
  GN-2004A-Q-19, GS-2004B-Q-31, GN-2004B-Q-16, GS-2005A-Q-11,
  GN-2005A-Q-11, GS-2005B-Q-6, GN-2005B-Q-7, GN-2006A-Q-7,
  GN-2006B-Q-10, and GN-2007A-Q-8.} \citep{how05,bro08}, the Very
Large Telescope (VLT)\footnote{Observations made with ESO Telescopes
  at the Paranal Observatory under program IDs 171.A-0486 and
  176.A-0589.} \citep{bal09}, and Keck\footnote{Some of the data
  presented herein were obtained at the W.M. Keck Observatory, which
  is operated as a scientific partnership among the California
  Institute of Technology, the University of California and the
  National Aeronautics and Space Administration. The Observatory was
  made possible by the generous financial support of the W.M. Keck
  Foundation.} \citep{ell08}.

\begin{figure}
\plotone{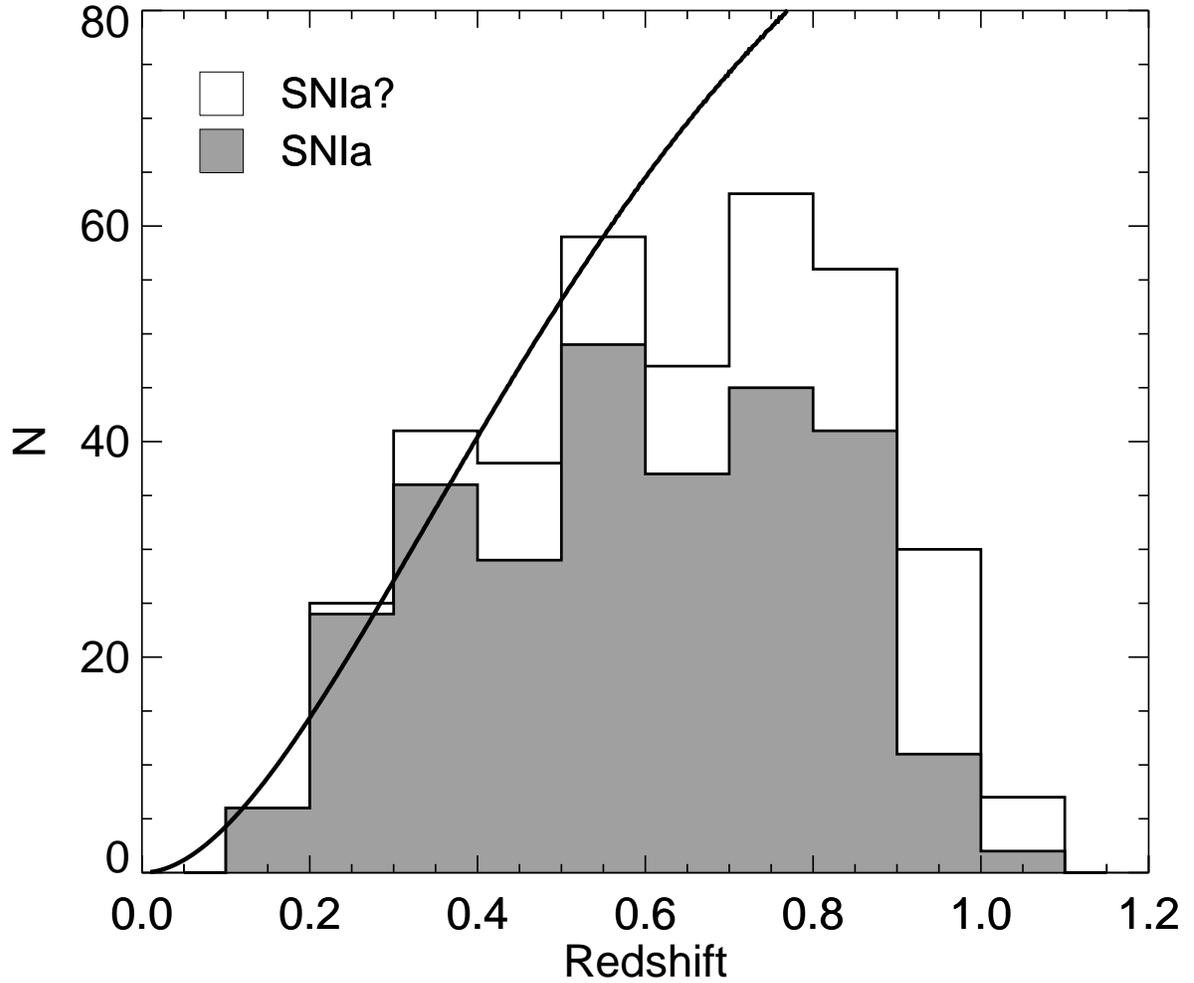}
\caption{The number of spectroscopic SNe~Ia from the SNLS as a
  function of spectroscopic redshift.  This plot includes confirmed
  SNe~Ia (shaded region) and probable SNe~Ia (open region) discovered
  up to the end of D3 observations in June 2007.  The solid curve
  represents the rise in number expected purely due to the increasing
  volume at greater redshifts, scaled to match the observed value in
  the $z=0.5-0.6$ bin.}
\label{fig:nz}
\end{figure}

The redshift distribution of the spectroscopically-identified SNe~Ia
found by SNLS is shown in Figure~\ref{fig:nz}.  The spectroscopic
sample is fairly complete out to $z\sim0.6$, at which point the
numbers begin to drop off as the amount of exposure time required per
candidate increases significantly.  This effect on the incompleteness
of the SNLS is examined later in \S\ref{sec:malm}.

\begin{figure}
\epsscale{0.7}
\plotone{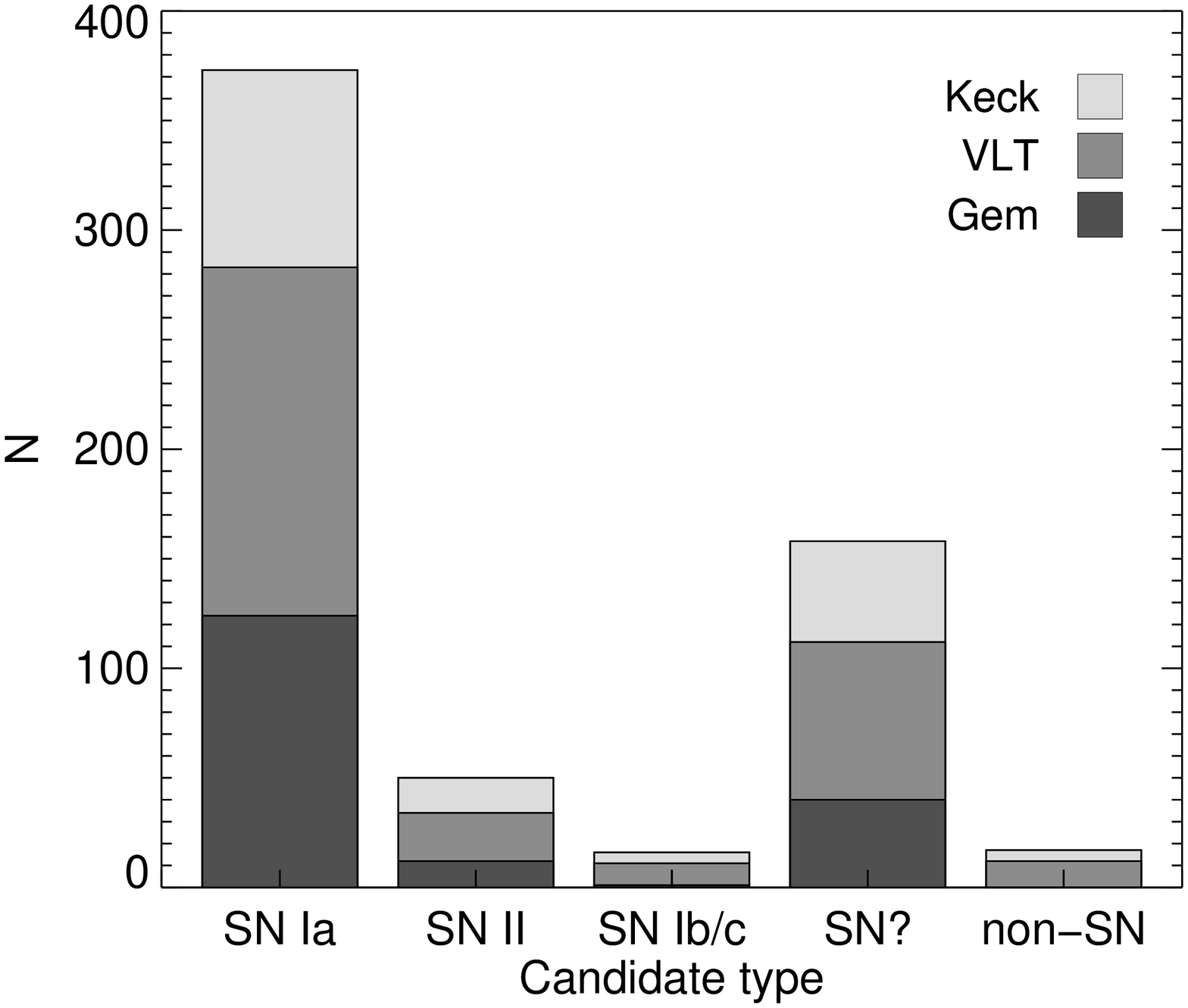}
\plotone{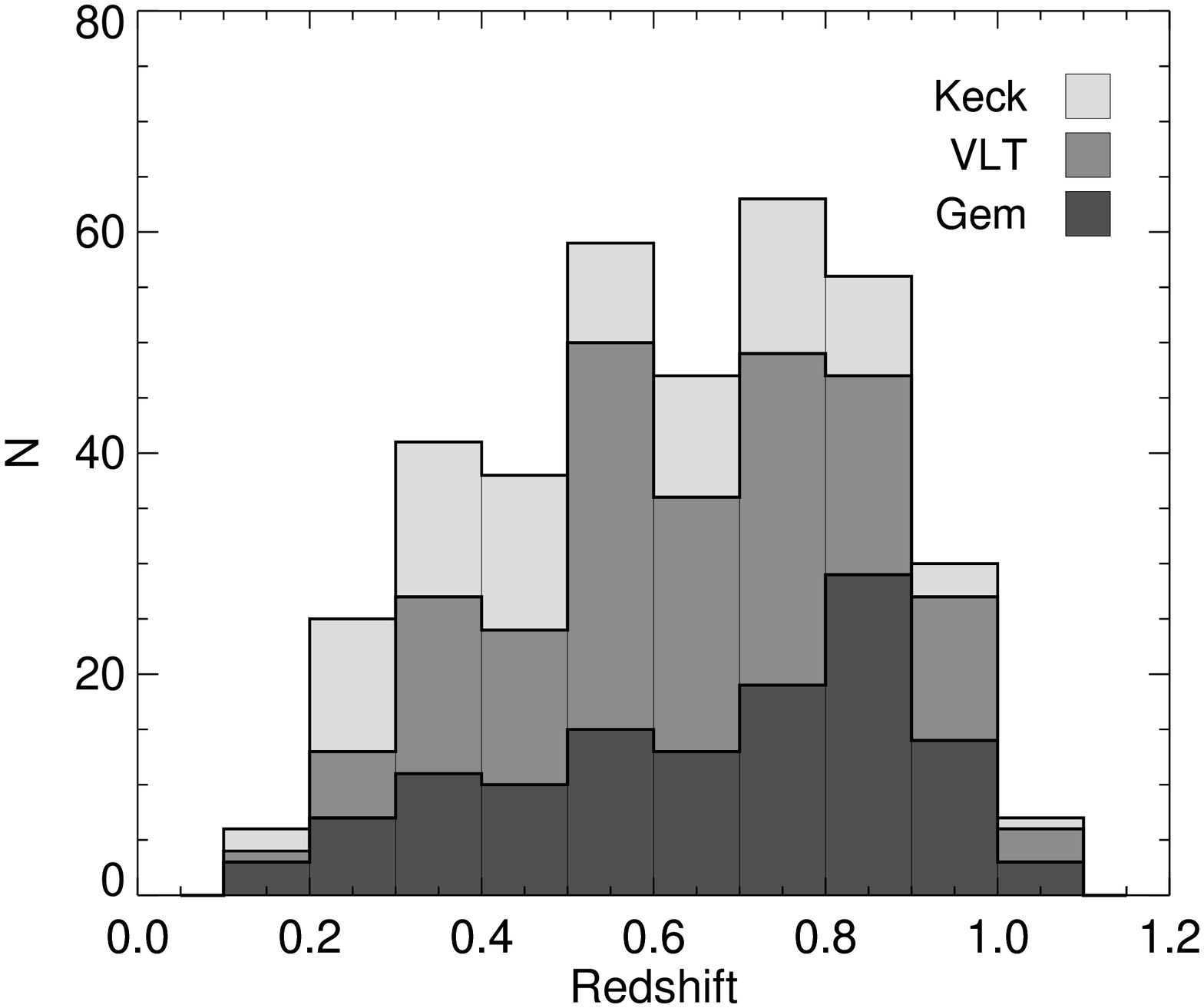}
\caption{Supernova candidates from the SNLS with spectroscopic
  follow-up observations.  {\it Top:} Candidate types as determined by
  spectroscopy from Gemini, VLT and Keck. {\it Bottom:} Redshift
  distribution for Type Ia SNe from the SNLS sample, as in
  Figure~\ref{fig:nz} but separated by telescope.  Both plots only
  include discoveries up to the end of D3 observations in June 2007.
  The SN~Ia category includes the probable (SN~Ia?) candidates.}
\label{fig:spec}
\epsscale{1}
\end{figure}

Histograms showing the breakdown of the sample by type and
spectroscopic redshift from the various telescopes are presented in
Figure~\ref{fig:spec}.  Candidates labeled as ``SN?'' are those that
have not been confirmed as SNe~Ia or core-collapse SNe; this category
may also include other types of variable objects that were not easily
distinguished from supernovae.  The non-SNe category mostly includes
active galactic nuclei (AGNs) and other objects, most of which were
observed prior to the implementation of the photometric fitting
technique for follow-up ranking described in \S\ref{sec:ranking}.

The median spectroscopic redshifts of the SNe~Ia observed at the three
observatories are $z=0.72$ (Gemini), $z=0.62$ (VLT), and $z=0.52$
(Keck).  Higher-redshift candidates were more frequently sent to
Gemini because of the larger observing overheads, and because the
nod-and-shuffle mode available at Gemini is useful for removing sky
lines to obtain better spectra of high-$z$ SNe. The Keck observations
included several SN~Ia programs with different science goals.

\subsection{Candidate numbers}

\begin{figure}
\plotone{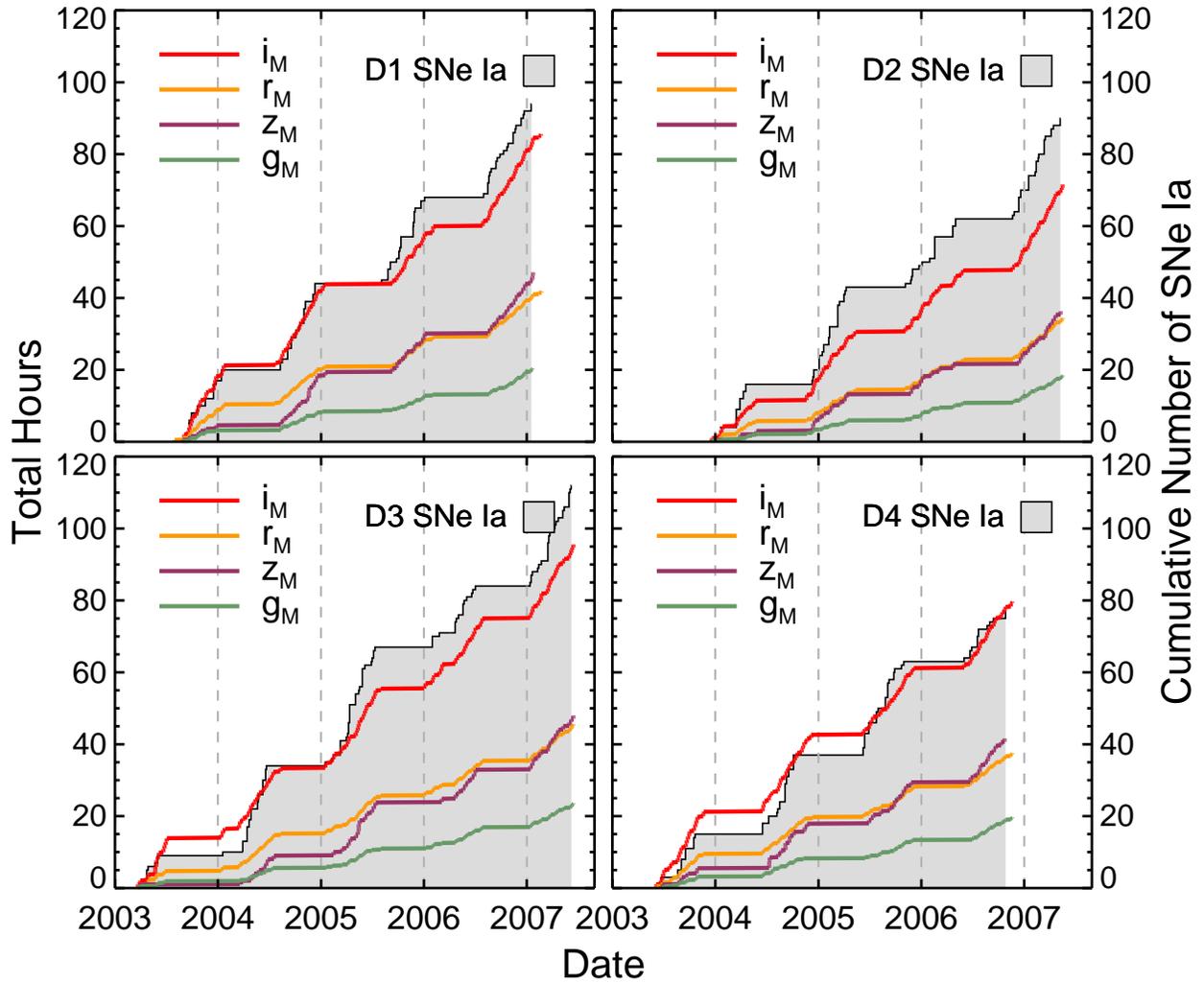}
\caption{A timeline showing the cumulative number of probable and
  confirmed SNe~Ia discovered in each Deep Field by SNLS (shaded
  area), along with the total exposure time obtained in the separate
  filters.}
\label{fig:exptime}
\end{figure}

In its first four years of full operation (to July, 2007)\footnote{In
  June 2007, the $i_M$ filter for MegaCam was destroyed during a
  malfunction of the filter jukebox.  Candidates discovered after this
  date were observed with a new $i_M$ filter, requiring new references
  and calibrations, and are still currently under analysis.
  Therefore, they are not included in the sample described in this
  paper.}, the SNLS discovered $>3000$ astrophysical transient events. Of
these, $373$ were spectroscopically identified as SNe~Ia.  This is the
same SNLS sample used in the SN~Ia rates analysis of \citet{per10}.
As expected, there is a strong correlation between the number of SN~Ia
detections and the total observing time in $i_M$, as demonstrated in
Figure~\ref{fig:exptime}.  This figure contains four plots, one per
field, each showing the running total of SNLS exposure time per filter
(with $i_M$ in red) for comparison with the cumulative number of
SNe~Ia discovered (the shaded areas).

\begin{deluxetable}{lcc}
\tablewidth{0pt}
 \tablecaption{SNLS candidates to 2007-06-30\label{tab:cand}}
\tablehead{
  \colhead{Type} & \colhead{$N$} & \colhead{$N_{\rm spec}$}
}
\startdata
SN~Ia(?)    &   373 &  373 \\
SN~II(?)    &    51 &   50 \\
SN~Ib/c(?)  &    17 &   17 \\
Other SN(?) &  1309 &  166 \\
non-SN      &  1360 &   22 \\ \hline
Total       &  3110 &  628 \enddata
\end{deluxetable}

A summary by type for all SNLS transient detections (excluding moving
objects and subtraction artifacts) is provided in
Table~\ref{tab:cand}.  $N$ refers to the number of objects of each
type in the SNLS real-time database and $N_{\mathrm spec}$ is the
number of those with available spectroscopy. Trailing question marks
``(?)'' in the type label in Column 1 indicate that the number
includes both confirmed and probable identifications of that type.
SNe~Ia?\ are classified as ``probable'' because their spectra best
match that of a Type~Ia, but another type of supernova --- typically a
SN~Ic --- cannot be conclusively ruled out from the spectrum alone.
SNe~Ia? are equivalent to the SNe~Ia$^\ast$ described in \citet{how05}
and \citet{bal09}. New SN-like detections were given the default label
``SN?'' in the database until they could be identified via
spectroscopy. (The one exception is a SN~II-P SN, which could be
identified from the obvious plateau in its light curve.)  Three of the
objects categorized as SN~Ib/c(?)\ are actually confirmed SNe~Ic:
SNLS-03D4aa, SNLS-04D4jv, and SNLS-06D4dv \citep{bal09}.

We believe the bulk of the ``Other SN'' category to be real SN events
that were not followed spectroscopically due to a lack of follow-up
time. Many of these are very faint (in some cases below the
spectroscopic limit) and likely high redshift; some had colors that
were clearly incompatible with SNe Ia. Others will be SN~Ia events in
our redshift range. These candidates will be analyzed further to
measure the volumetric SN~Ia rate in \citet{per10}. The ``non-SN''
category includes a large number of other astrophysical transient
events, including AGN and variable stars, identified via their erratic
light curves and central location in their ``hosts''.

We emphasize that the relative numbers of the different types of SNe
in Table~\ref{tab:cand} are strongly affected by spectroscopic
selection procedures.  The total numbers are also influenced by
factors such as losses due to poor winter weather conditions on Mauna
Kea, as well as the slow ramp-up of observing during the pre-survey
period.  These observational effects are incorporated into the
candidate recovery statistics that are measured in the next section.

\section{SNLS Detection Efficiency}
\label{sec:comp}

A thorough understanding of the recovery statistics in the RTA
pipeline provides the basis for calculating sample selection biases.
Detailed Monte Carlo simulations were carried out to measure the
efficiency of SN~Ia recovery from the SNLS real-time images. Although
these simulations could have been performed concurrently with the
real-time candidate searching \citep[e.g.,][]{dil08}, in the SNLS we
perform these simulations completely separately from the daily search
activities.  A description of the input test population is provided in
\S\ref{sec:sims}, and the results of the simulations are outlined
afterwards in \S\ref{sec:simresults}.

\subsection{SNe~Ia input parameters}
\label{sec:sims}

In all, 15 realizations of 10\,000 artificial SNe each were performed
individually for every Deep Field observing season up to the end of D3
in the summer of 2007.  A ``field-season'' represents the period over
a year during which a given field was visible: in this current sample,
D3 was visible for five seasons (counting the pre-survey observing),
while the rest of the fields were each visible for four.  This yields
a total of more than $2.5 \times10^6$ artificial SNe tested in the
simulations during 17 field-seasons.

To begin, a random redshift was selected for each input artificial SN
from a uniform distribution over $0.1\leq z \leq 1.2$.  The artificial
SNe were then randomly assigned to hosts chosen from the SNLS field
galaxy catalogs compiled from measurements of deep image stacks
\citep[see][]{sul06b}.  The limiting magnitude of the reference stacks
is $i_M\sim25.7$~mag (AB), much fainter than the detection limit for
any given epoch.  Photometric redshifts for the field galaxies were
calculated based on the P\'EGASE.2 galaxy spectral evolution code
\citep{leb04}.  A random host galaxy within $z\pm0.02$ of the selected
redshift was picked in the appropriate Deep Field, weighted by a
probability approximated by the two-component model for SN~Ia rates
(SNR):
\begin{equation}
\label{eq:aplusb}
P(z) \sim \mathrm{SNR}(z) = A\times M(z) + B\times\mathrm{SFR}(z)
\end{equation}
\citep{man05,sb05}.  $A=5.3\Aunit$ is the mass coefficient, and
$B=3.9\Bunit$ is the coefficient of the star-formation rate (SFR) term
\citep[adopted from][]{sul06a}. As a precaution against overcrowding,
a field galaxy was not permitted to host more than one artificial SN
during any given season.

The artificial SNe were assigned galactocentric positions drawn from a
2D Gaussian distribution about the host centroid\footnote{The galaxy
  light in the stacks has already been convolved with the seeing
  ($\sim 0.5\arcsec$), yet the SN positions are accurate to $\sim
  0.1\arcsec$.  This will slightly underestimate the true central
  concentration of the input SN distribution. However, we find that
  the recovery fraction is quite flat with radius, as shown later in
  Figure~\ref{fig:comp_dmgal}.  This is also in agreement with the
  conclusions of \citet{how00}, who demonstrate that selection bias at
  small host radii is not significant for CCD observations of high-$z$
  SNe~Ia.}, verifying that each fell within the galaxy's detection
ellipse returned by SExtractor.  This produced a list of input
coordinates for the input sample according to the astrometric solution
of the reference stack.

The input objects were then randomly allocated stretch values drawn
from a uniform distribution in the range $0.5\leq s \leq 1.3$.  This
wide range was adopted in order to over-sample the parameter space of
interest, to include recovery statistics for potential sub-luminous
SNe in the SNLS data \citep{gon10}.  This population could later be
resampled to investigate the observed distribution
(\S\ref{sec:simresults}).

Each candidate was assigned a random peak date within the range
$\min(\mathrm{MJD}_\mathrm{obs})-20 \leq \mathrm{MJD}_\mathrm{peak}
\leq \max(\mathrm{MJD}_\mathrm{obs})+10$, where MJD$_\mathrm{obs}$ are
the observed dates of all of the SNLS images obtained for the field
and year being tested.  The $^{+10}_{-20}$ values are included to
account for candidates that were detected while fading at the start or
rising at the end of a field's observing period.  Observational culls
based on a required minimum number of observations before and after
peak could be applied later as needed to test for selection bias.

The peak $B$ magnitude appropriate for a Type~Ia at each chosen
redshift was calculated by incorporating the stretch-luminosity
relationship and an absolute magnitude derived from cosmology fits
assuming H$_0=70$.  The adopted cosmological parameters are
$\Omega_m=0.27$, $\Omega_\lambda=0.73$, and $\Omega_k=0$.  A random
dispersion in the peak $B$-band magnitude ($\Delta m_B = \Delta$mag)
was then applied after the stretch correction, drawing from a Gaussian
distribution with an intrinsic scatter of $\sigma_\mathrm{int}=0.15$.
Again, a broad dispersion was adopted so that the output distribution
could be resampled as needed.

A rest-frame spectrum was extracted from an SN~Ia spectral template
\citep{hsi07} and was scaled up or down to the appropriate $B$
magnitude for each artificial object (with dispersion applied), thereby
also affecting $U$ and $V$.  The output peak $B$ magnitude was found
by integrating over the Landolt filter functions.  The $\ub$ and $\bv$
colors at peak were calculated using a bilinear model of the
stretch-color relationship with a break at $s=0.8$ \citep{gar04}, such
that:
\begin{equation}
\label{eq:stretch}
\bv =
\begin{cases}
-0.039 & (s \geq 0.8) \\
1.711 - 2.187\,s & (s < 0.8)
\end{cases}
\end{equation}
This $s-c$ relationship model was derived from fits of color as a
function of stretch for the sample of low-$z$ SNe~Ia and the SNLS
spectroscopic sample at $z<0.6$ \citep{gon10}.

This procedure yields $U$ and $V$ rest magnitudes, and thus raw
colors, for the artificial SNe.  Gaussian noise with a dispersion of
$\sigma=0.04$ was added independently to $U$ and $V$ to adjust the
colors, and the spectrum was ``mangled'' (i.e., color-corrected) to
those values \citep[see][]{hsi07}.

The spectral energy distribution was next reddened to account for
Milky Way extinction, $E(\bv)_\mathrm{MW}$ with $R_\mathrm{v,MW}=3.1$,
by randomly selecting from a one-sided (always positive) Gaussian
distribution with $\sigma=0.015$. Host extinction,
$E(\bv)_\mathrm{host}$, drawn from a uniform color distribution
within the range $-0.25 \leq c \leq 0.5$, was also incorporated by assuming
$R_\mathrm{v,host}=\beta-1.0$, with $\beta=3.0$ to match the SNLS
spectroscopic sample at low redshift.  Note that this does not
precisely match our current understanding of the SN~Ia color-color
relationship \citep{guy07, con08}, which does not entirely match
any measured dust law.  However, this has a negligible effect on
our estimates of Malmquist bias or SN rates as discussed later.

Finally, realistic SN~Ia light curves were generated using a similar
process as described above for calculating the peak magnitudes and
colors.  Randomly selected, isolated PSF stars in each object's CCD
(brighter than the object to add) were used as the artificial SN
models. Poisson noise was added to account for image gain. The stellar
models were scaled and inserted with their appropriate phases,
magnitudes, and positions into every pre-processed $i_M$ image
obtained by SNLS.  Due to the transient nature of SNe~Ia, most of the
input objects added to any particular image were fainter than the
detection limit, and no overcrowding effects were observed in the
trials.  The images were then processed and analyzed using the same
RTA procedure as described in \S\ref{sec:pipeline}. The output
detection lists for each epoch were compared with the input records to
calculate the recovery efficiencies as a function of several supernova
parameters.

The strength of this straightforward yet rigorous approach is that it
makes use of {\em all of the actual images obtained} by SNLS to
measure recovery rates.  This avoids the need to make assumptions
about the detailed frequency and spacing of the observations,
technical losses (e.g., missing CCDs), weather conditions, other
observational effects, or to use simplified parameterizations to
represent the recovery efficiencies.

Before calculating the recovery statistics, the input distribution of
artificial SNe~Ia is first resampled in $z$, $s$, $c$, and
$\Delta$mag.  The goal is to represent the underlying population of
(non-peculiar) SNe before selection effects by adopting distributions
of light-curve width, color, and $\sigma_{\mathrm{int}}$ that match
the observed distributions as a function of redshift in the selected
sample.  The low-redshift portion of the sample ($z \la 0.6$) is
extremely useful for constraining these distributions, but note that
we are making the assumption that the underlying demographics of the
population do not change significantly more strongly than the $A+B$
model of Eq.~\ref{eq:aplusb} suggests.

Rather than re-sample the input distributions, we instead simply
weight the distributions to obtain the same effect while obtaining
higher statistics in our measurement by not throwing out any objects.
We weight the SN distribution to $\sigma_{\mathrm{int}} = 0.11$ mag
based on the cosmological fits in \citet{guy10, con10}.  This differs
from the value adopted therein because the simulated SNe used here do
not include uncertainties in the input light-curve model or other
systematics which tend to increase the measurement uncertainties.  We
find that the underlying stretch distribution is well described by a
two-component Gaussian distribution.  A two-component Gaussian could
not provide a good match to the observed color distribution and the
evolution in the observed mean color with redshift due to selection
effects. Instead, we have adopted the convolution of a exponential
distribution and a Gaussian, similar in form to that of \citet{jha07}.

\begin{figure} 
\plottwo{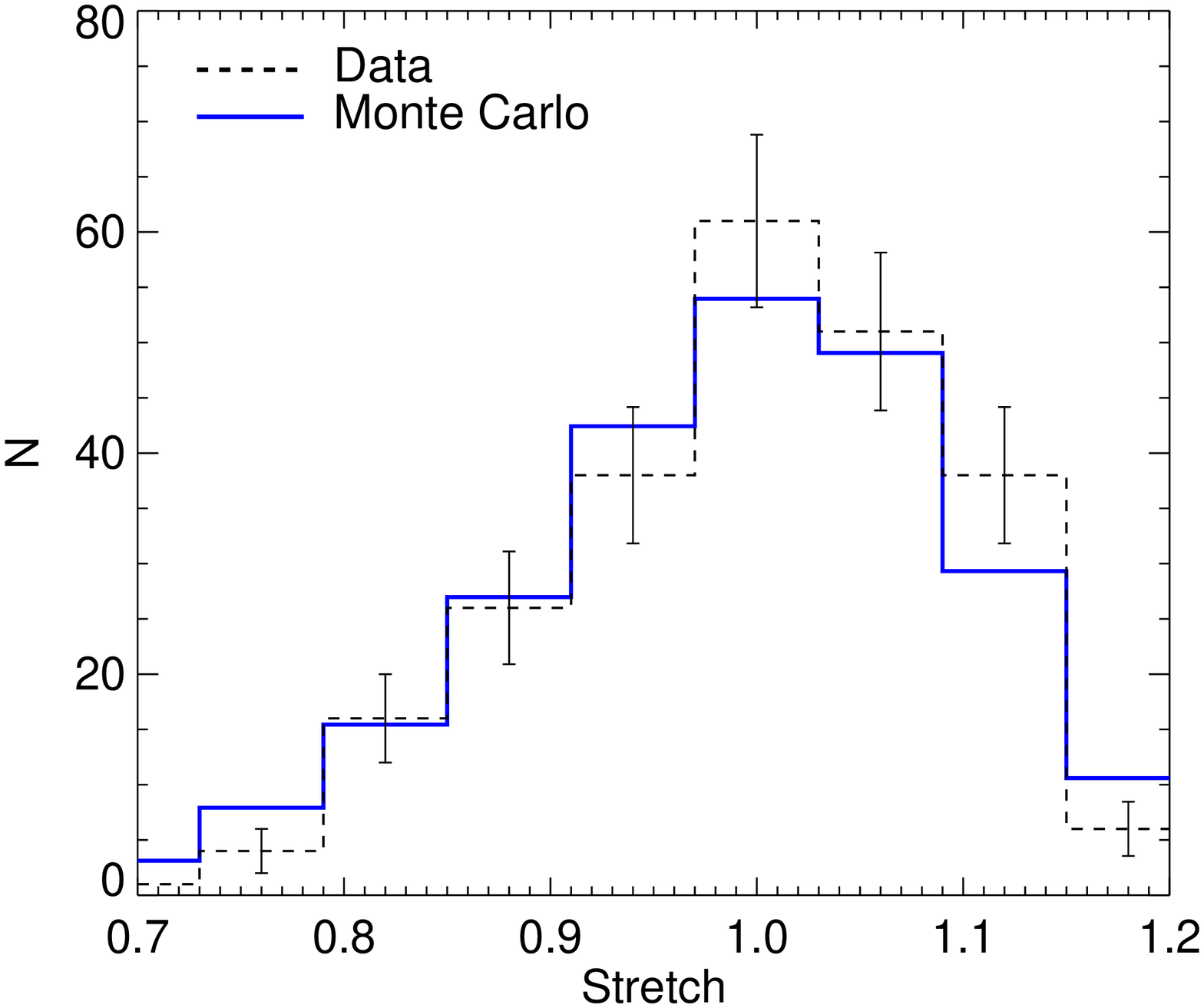}{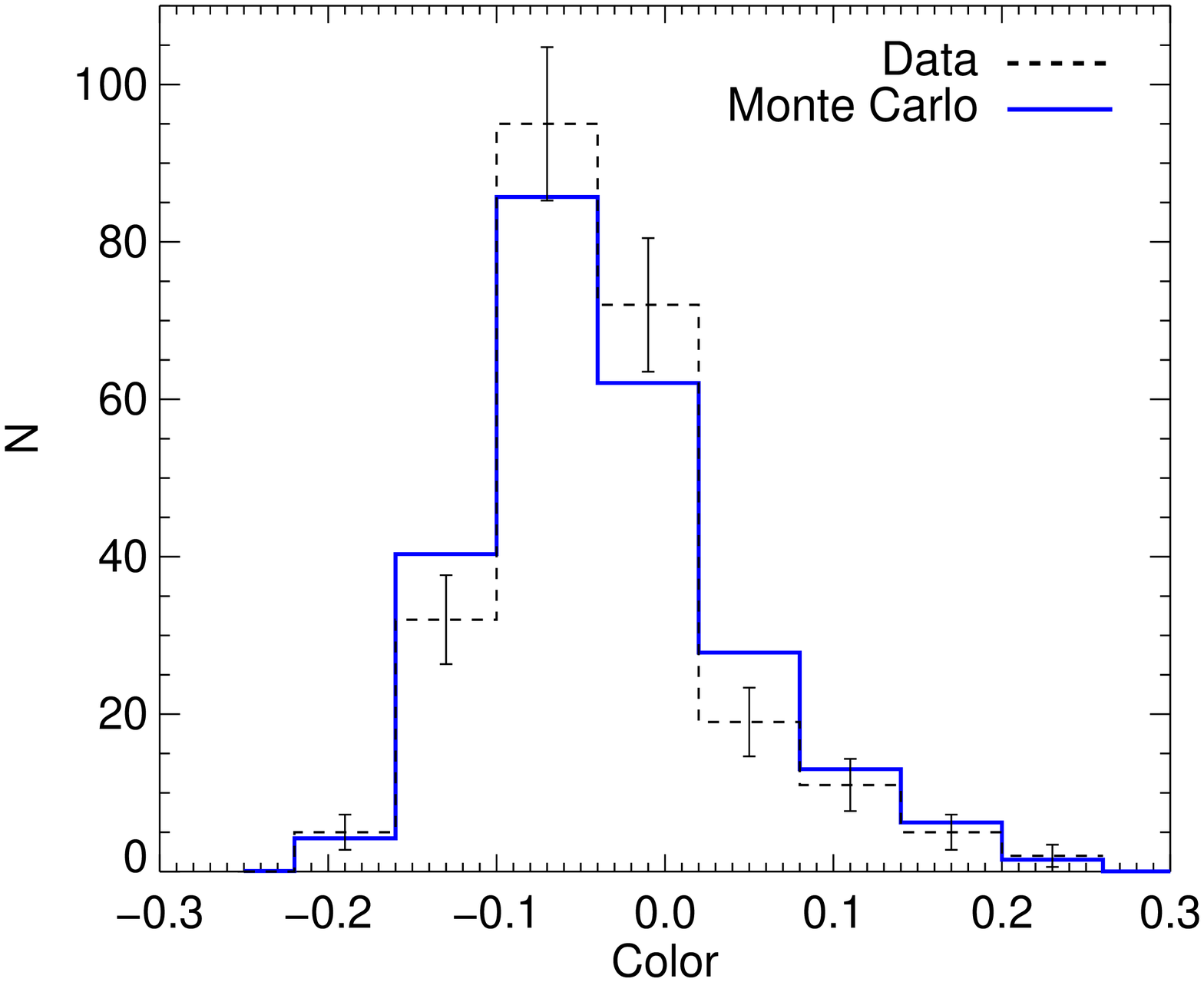}
\caption{The recovered stretch (left) and color (right) distributions
  compared with those observed for the 3rd-year SNLS cosmological
  sample \citep{guy10}.}
\label{fig:scolorcompare}
\end{figure}

\begin{figure} 
\plotone{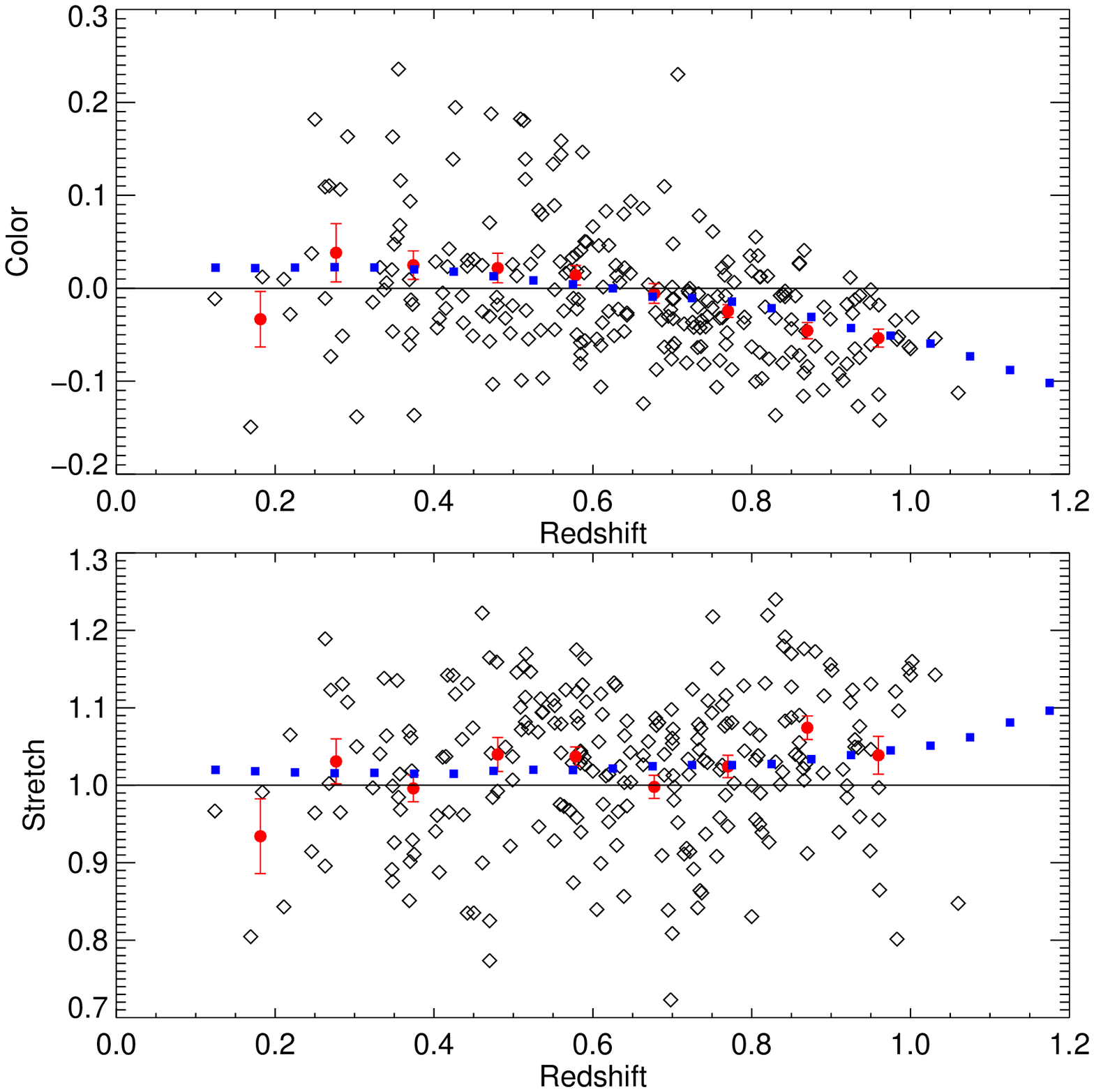}
\caption{The recovered stretch evolution in the mean stretch and color
  (blue squares) compared with the actual stretch and color
  distributions (averages shown as red circles) as a function of
  redshift.}
\label{fig:propevolve}
\end{figure}

The parameters of the stretch model are (mean, $\sigma$, peak value)
$0.92, 0.11, 0.37$ for the first Gaussian and $1.06, 0.08, 0.80$ for
the second.  For the color model, the $\sigma$ of the Gaussian is
0.04, the $\tau$ of the exponential distribution $0.155$, and the mean
value of the distribution 0.07.  Note that we do not interpret the
color distribution as arising from host-galaxy extinction, but simply
use this model as a useful empirical parameterization.  The resulting
color and stretch distributions are compared with the actual 3rd-year
cosmological sample in Figure~\ref{fig:scolorcompare}, and the
evolution of the mean parameters with redshift are shown in
Figure~\ref{fig:propevolve}.

\subsection{Recovery results}
\label{sec:simresults}

\begin{figure} 
\epsscale{0.65}
\plotone{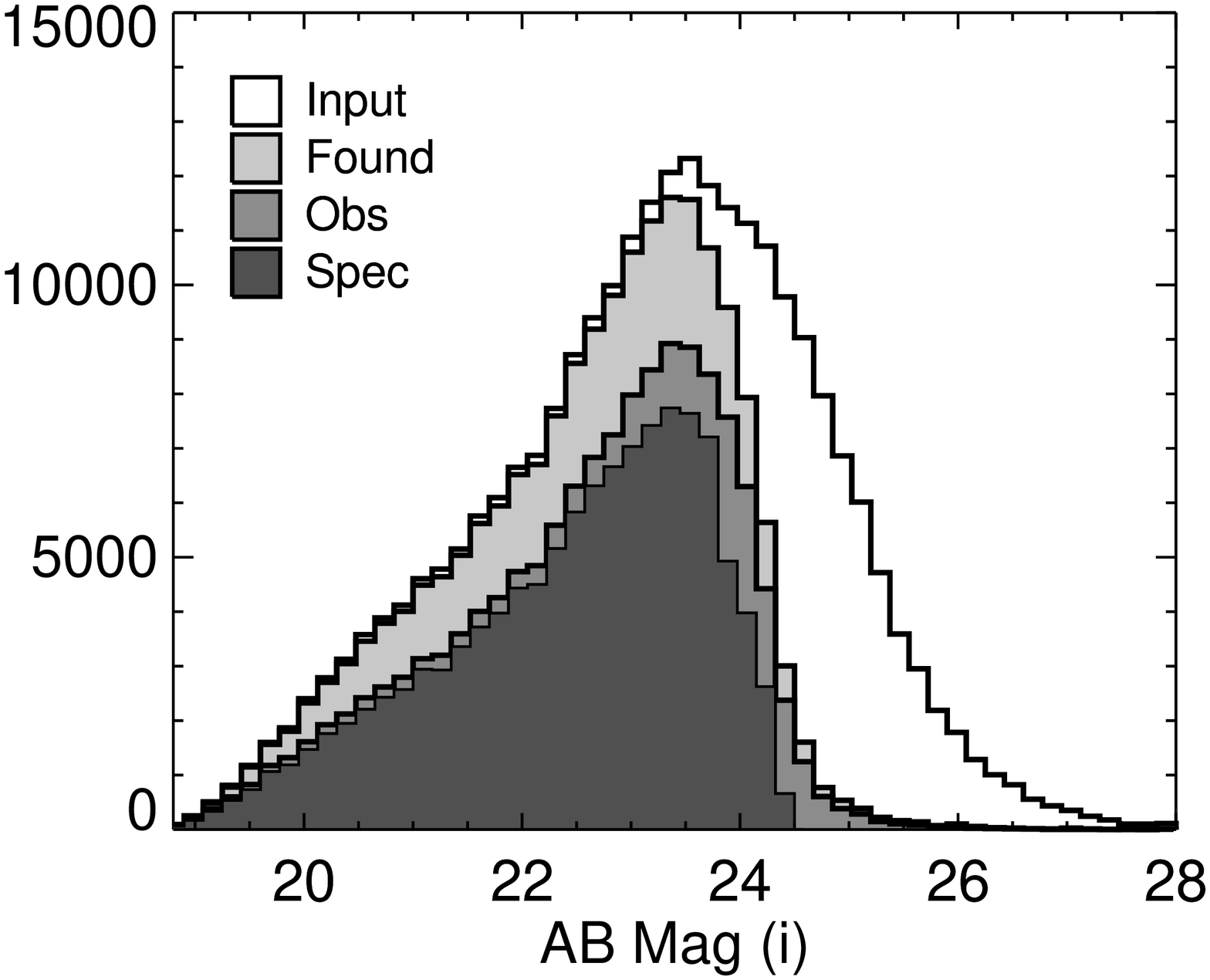}
\plotone{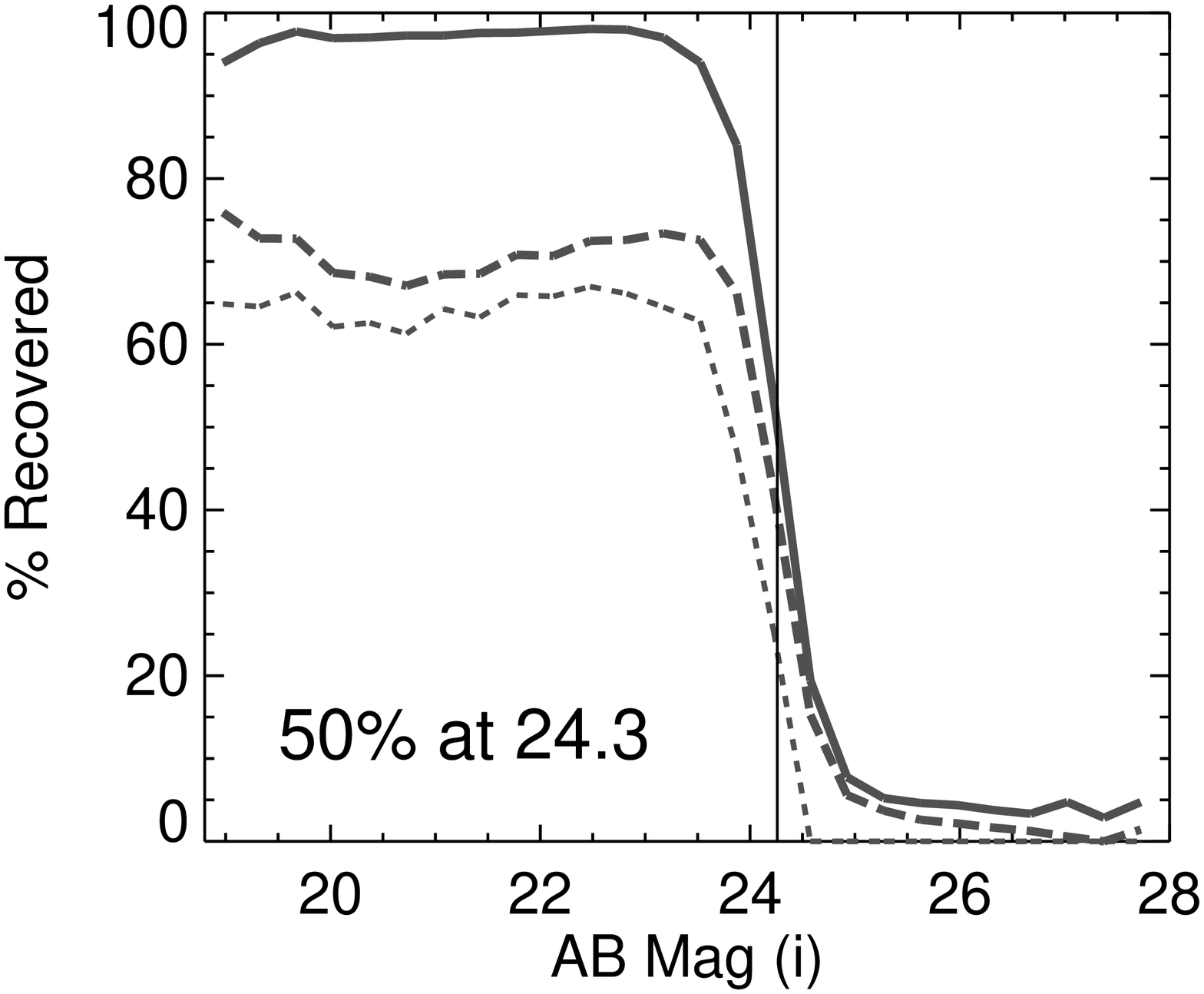}
\caption{The upper panel shows the input (unfilled) and recovered
  (light grey) distributions in peak $i_M$ magnitude for the
  artificial SNe~Ia used in the recovery tests.  The effects of adding
  the observational and spectroscopic constraints are shown by medium
  and darkest grey histograms, respectively.  The lower panels present
  the recovered fraction for the found sample (solid line) and with
  the observational and spectroscopic cuts (dashed and dotted lines,
  respectively).  The $50\%$ incompleteness limit for the real-time
  analysis lies at $i_\mathrm{AB}=24.3$~mag (AB), represented by the
  vertical line in the lower diagram.}
\label{fig:comp_mag}
\epsscale{1}
\end{figure}

\begin{figure}
\includegraphics[width=2.1in]{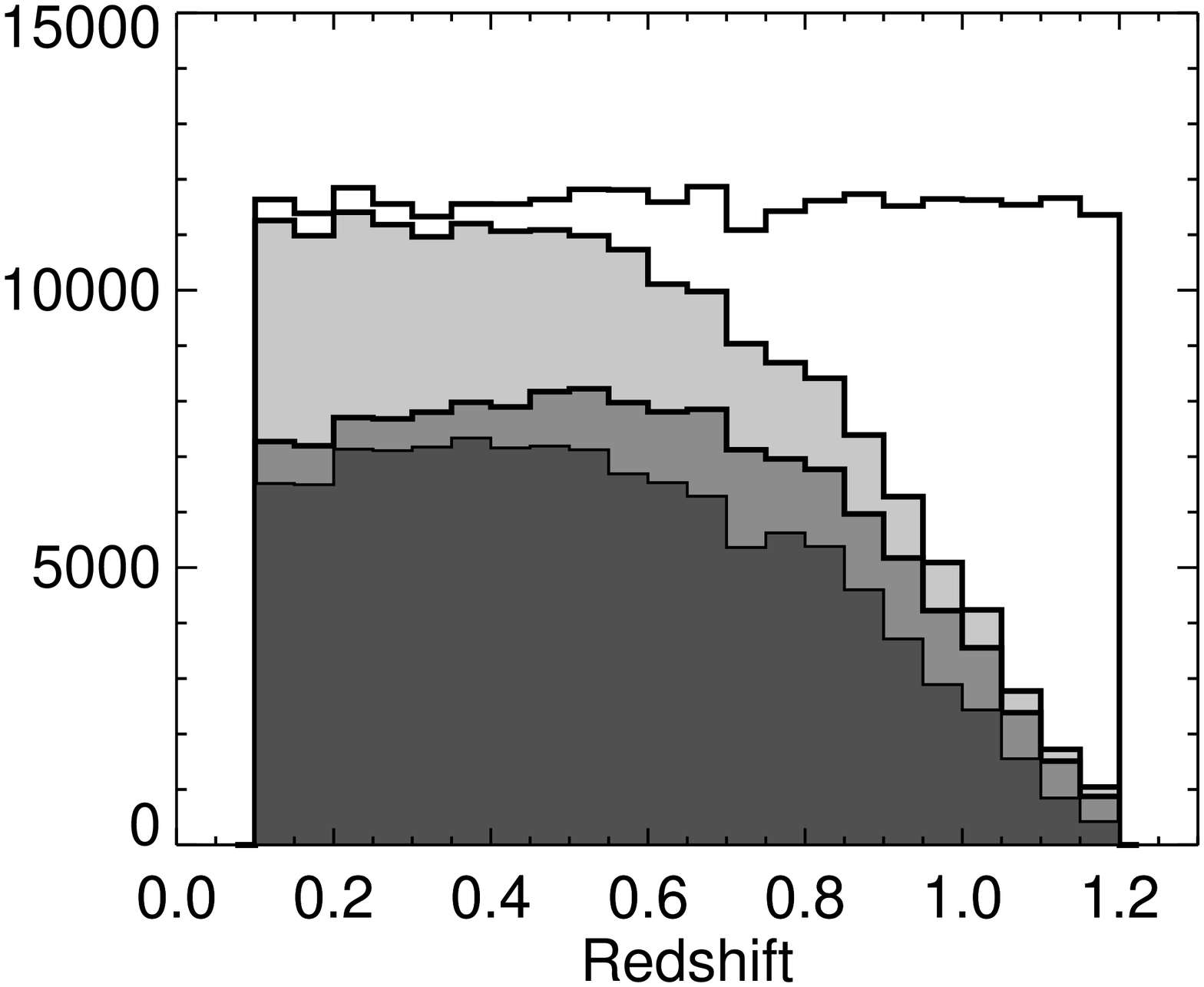}
\includegraphics[width=2.1in]{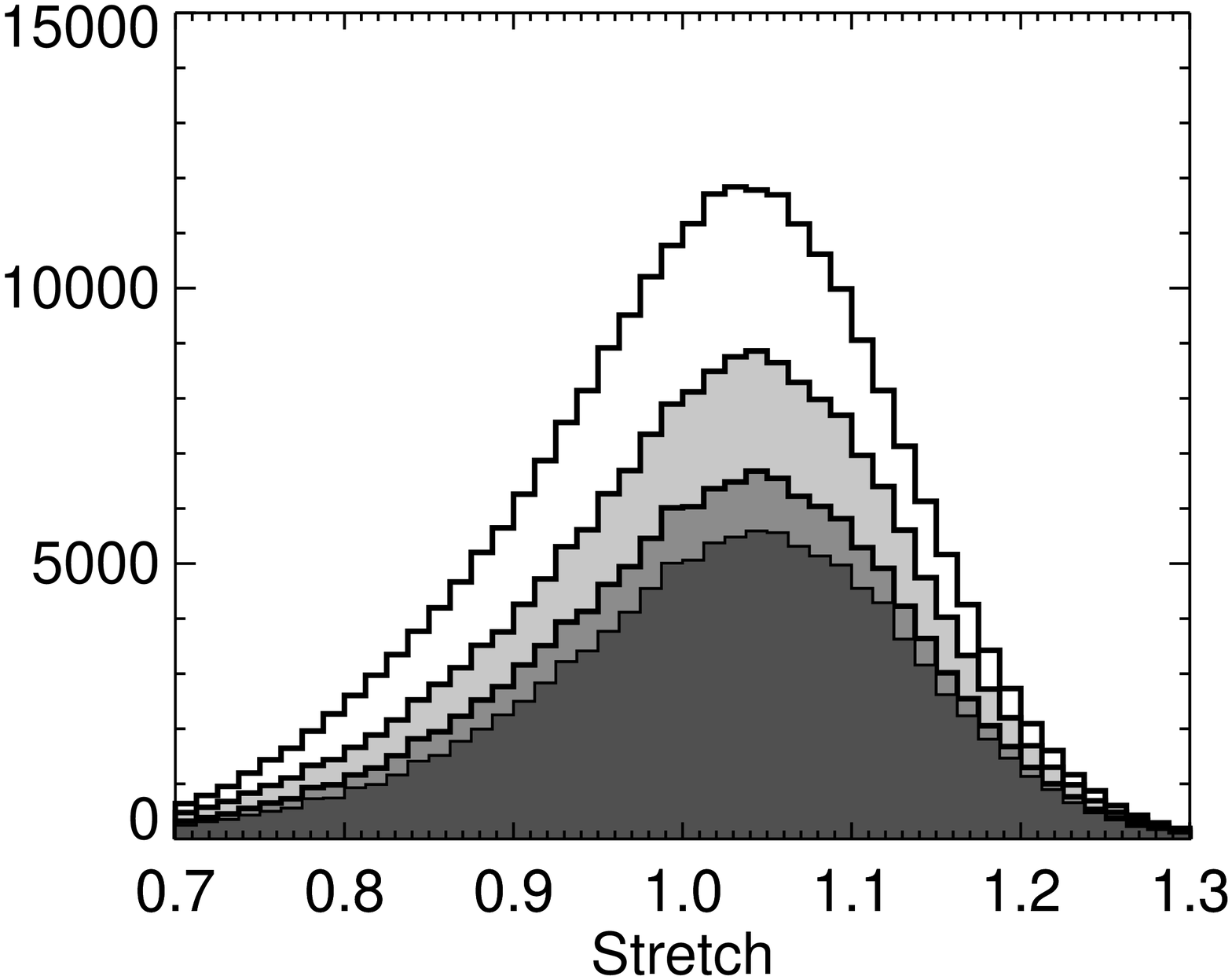}
\includegraphics[width=2.1in]{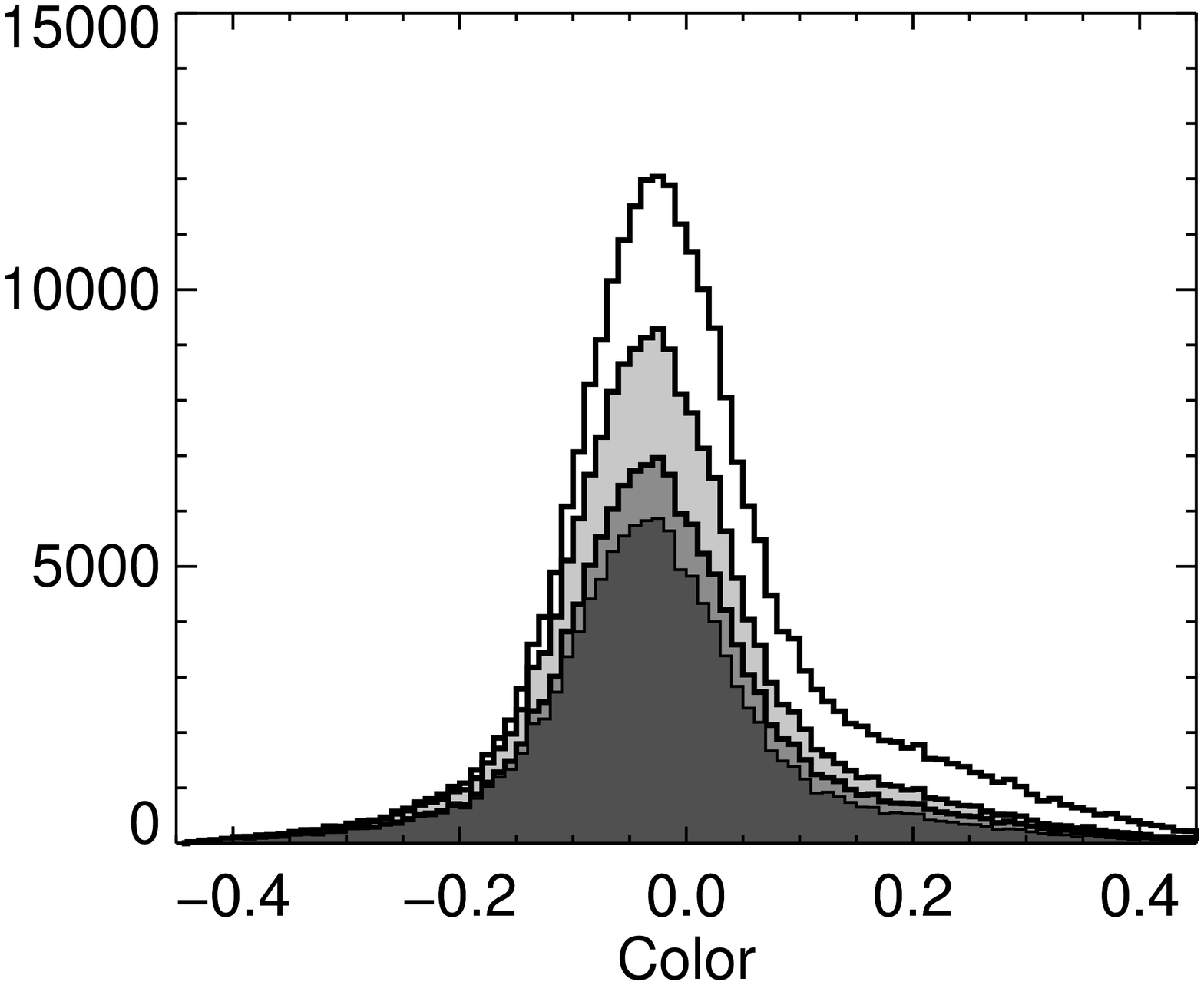}\\
\includegraphics[width=2.1in]{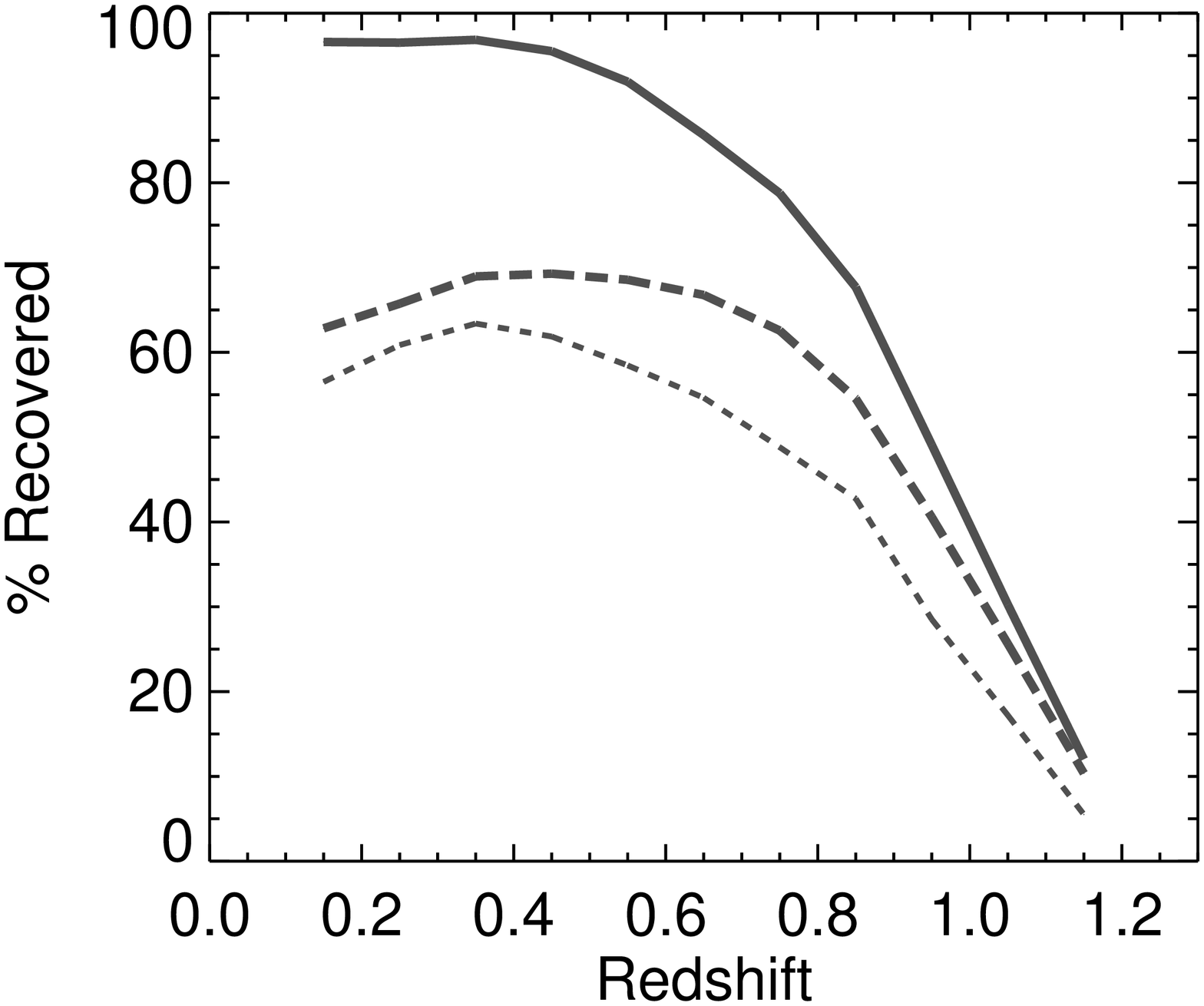}
\includegraphics[width=2.1in]{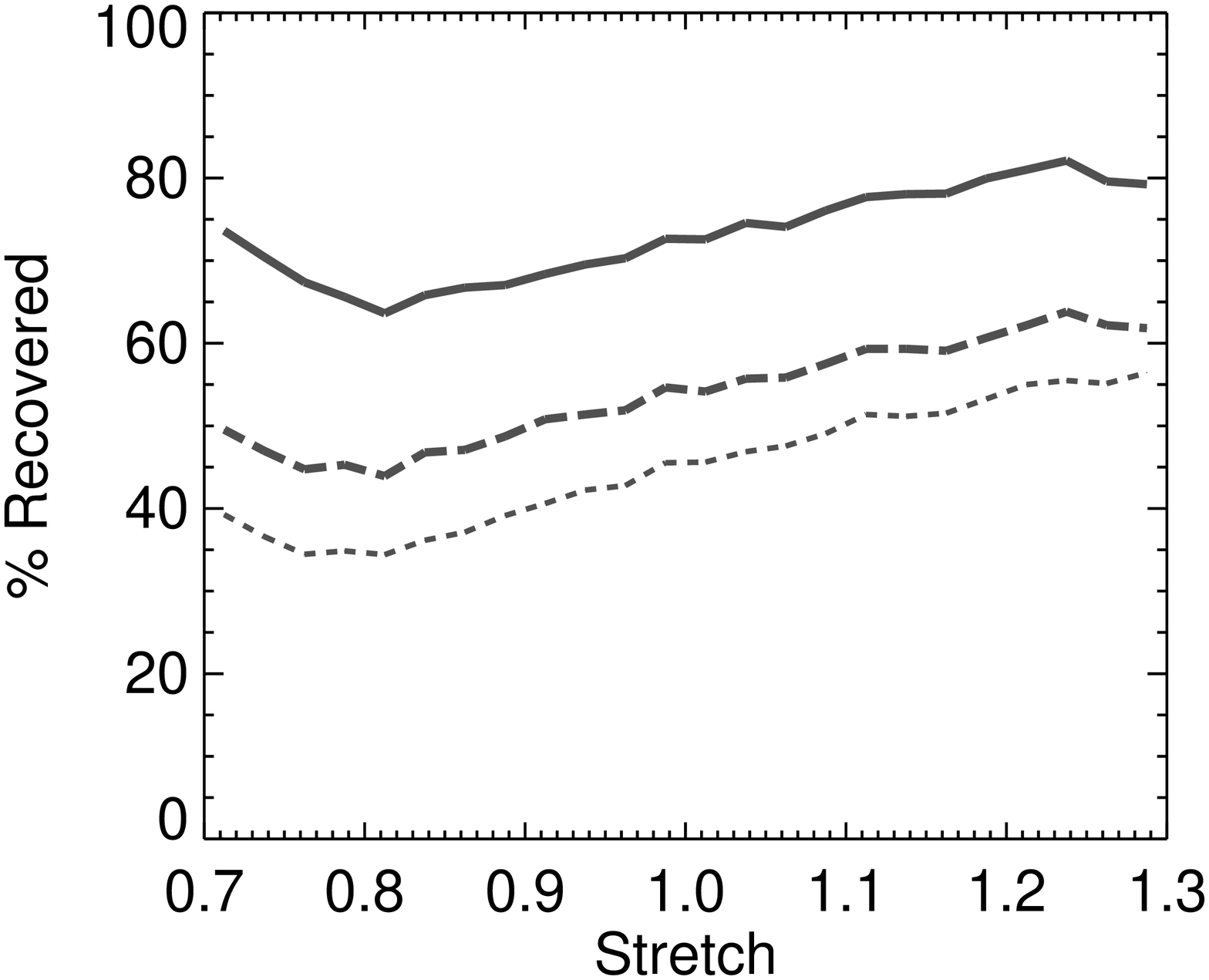}
\includegraphics[width=2.1in]{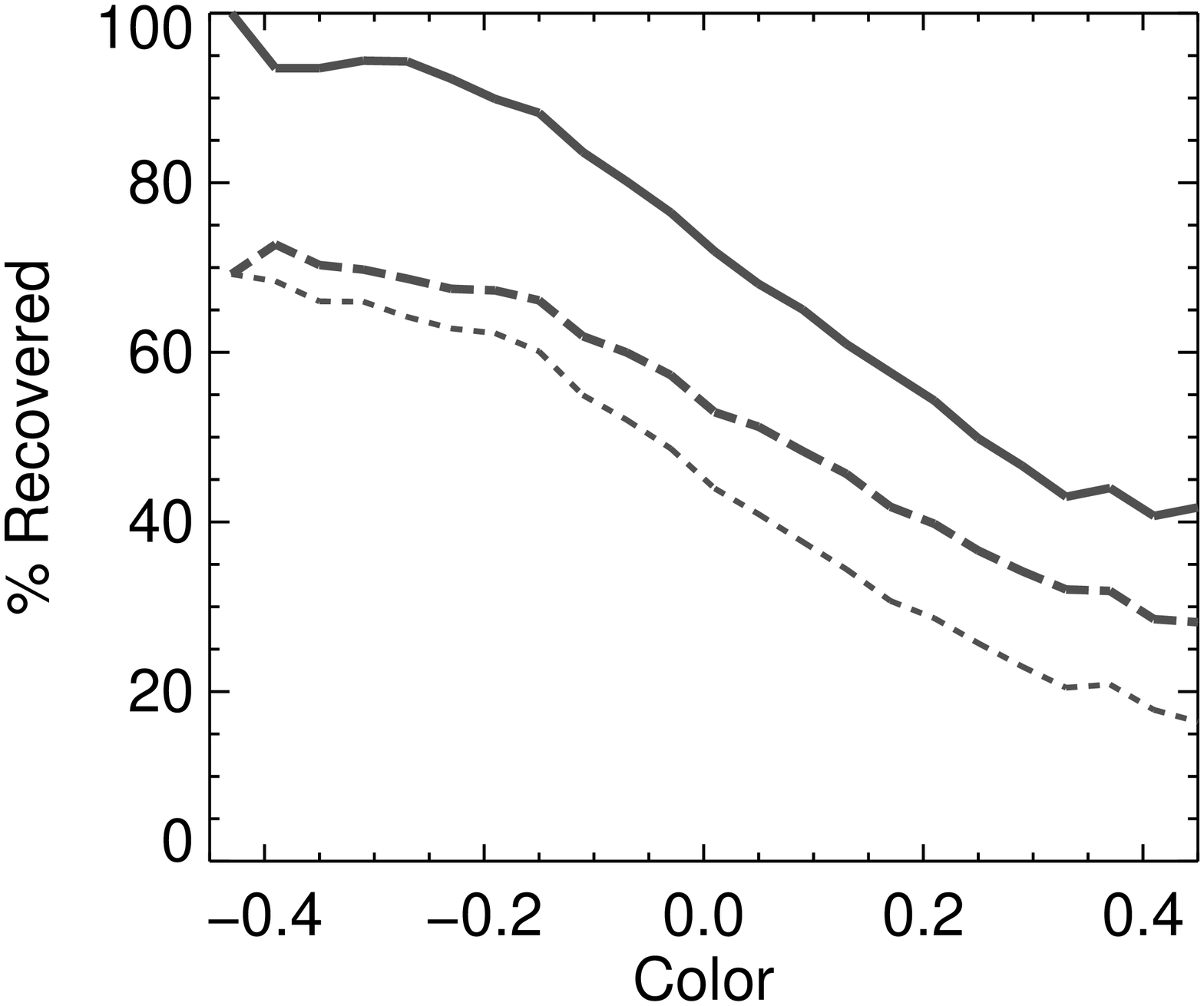}
\caption{SN~Ia recovery as a function of redshift (left), stretch
  (center), and color (right).  The histograms in the upper panels
  show the input distributions (unfilled) and the recovered sample
  (light grey), along with the addition of the observational and
  spectroscopic cuts (medium and darkest grey histograms,
  respectively).  The lower panels present the recovered fraction for
  the found sample (solid line) and with the observational and
  spectroscopic cuts (dashed and dotted lines, respectively).}
\label{fig:comp_zsc}
\end{figure}

\begin{figure}
\includegraphics[width=2.1in]{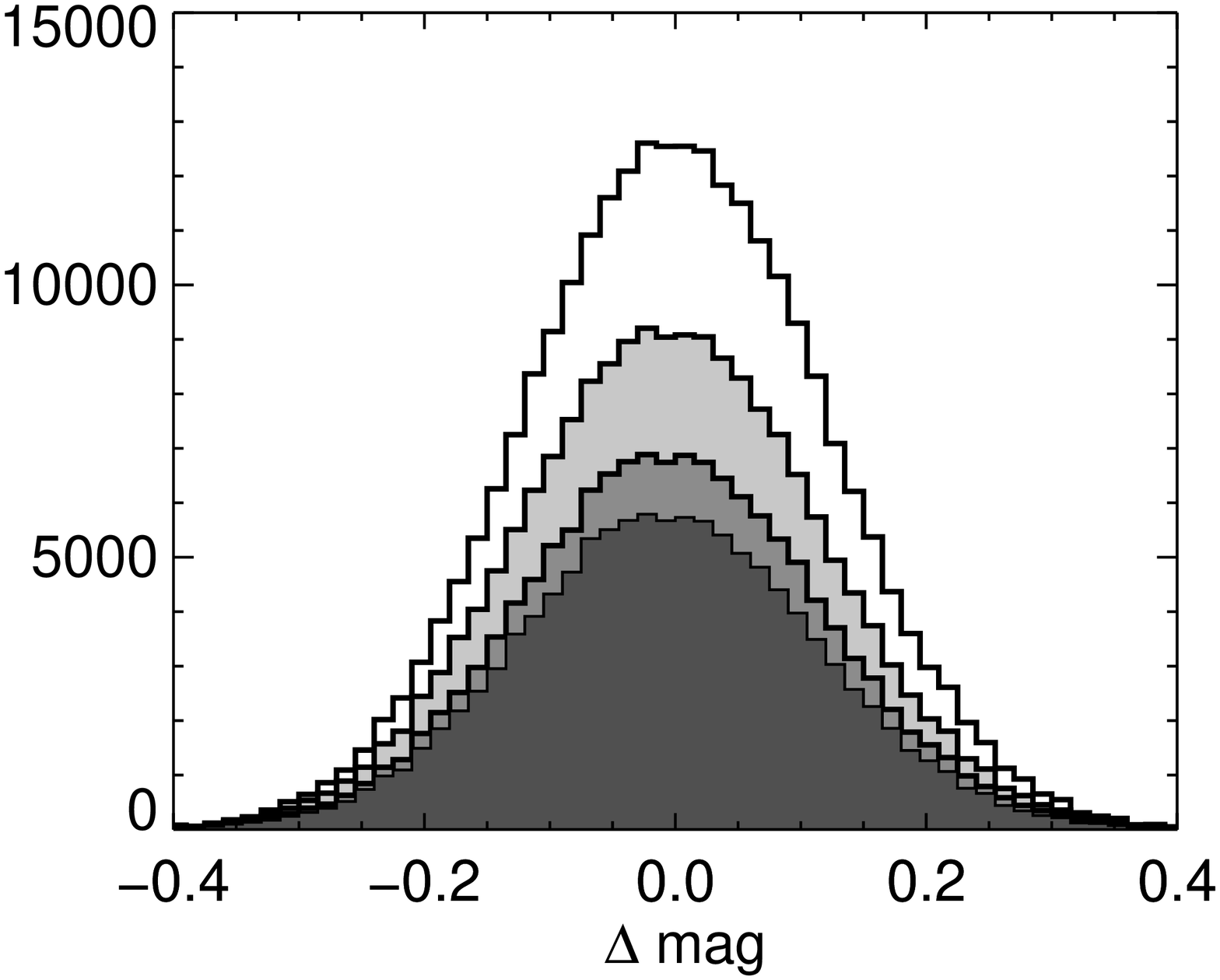}
\includegraphics[width=2.1in]{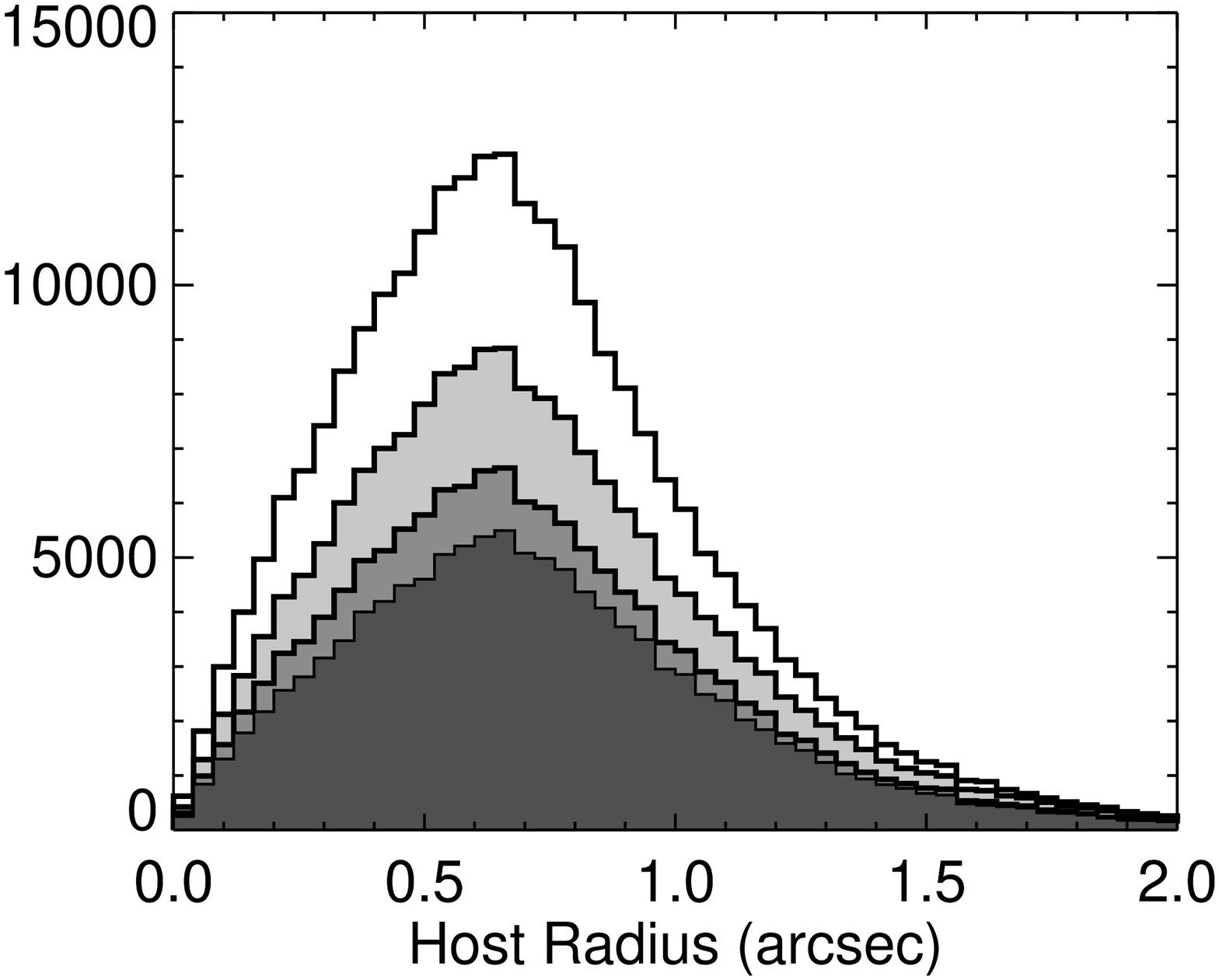}
\includegraphics[width=2.1in]{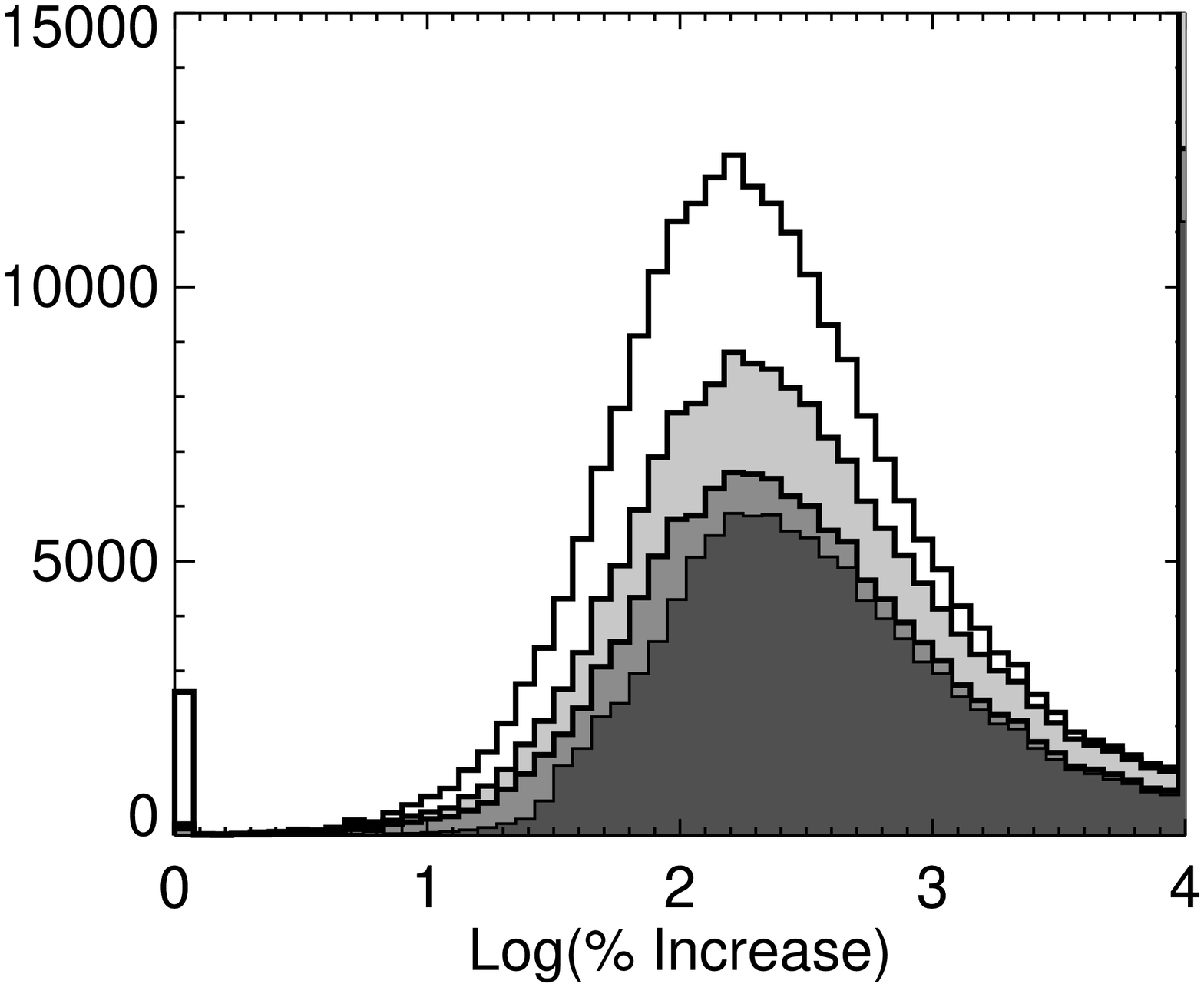}\\
\includegraphics[width=2.1in]{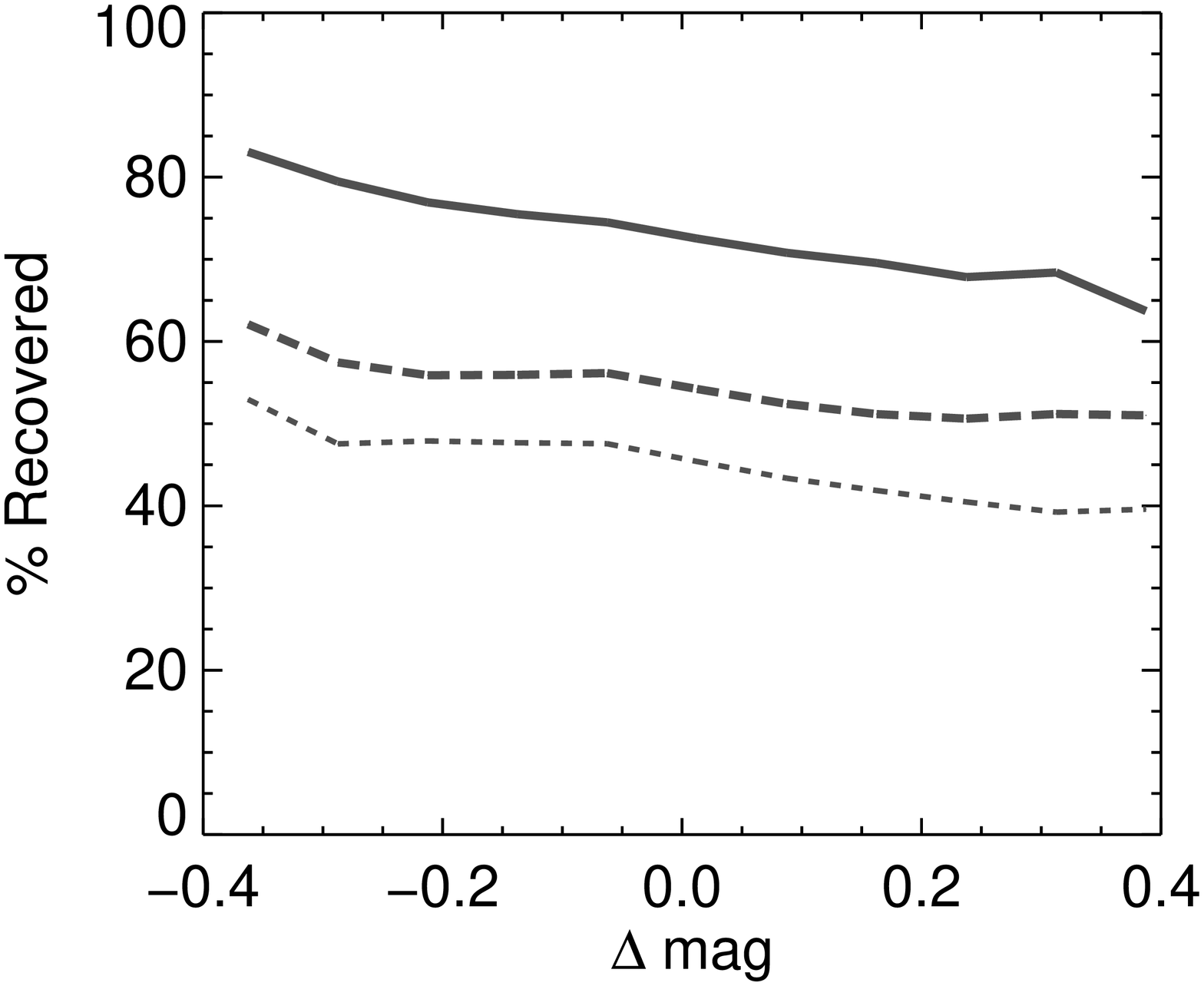}
\includegraphics[width=2.1in]{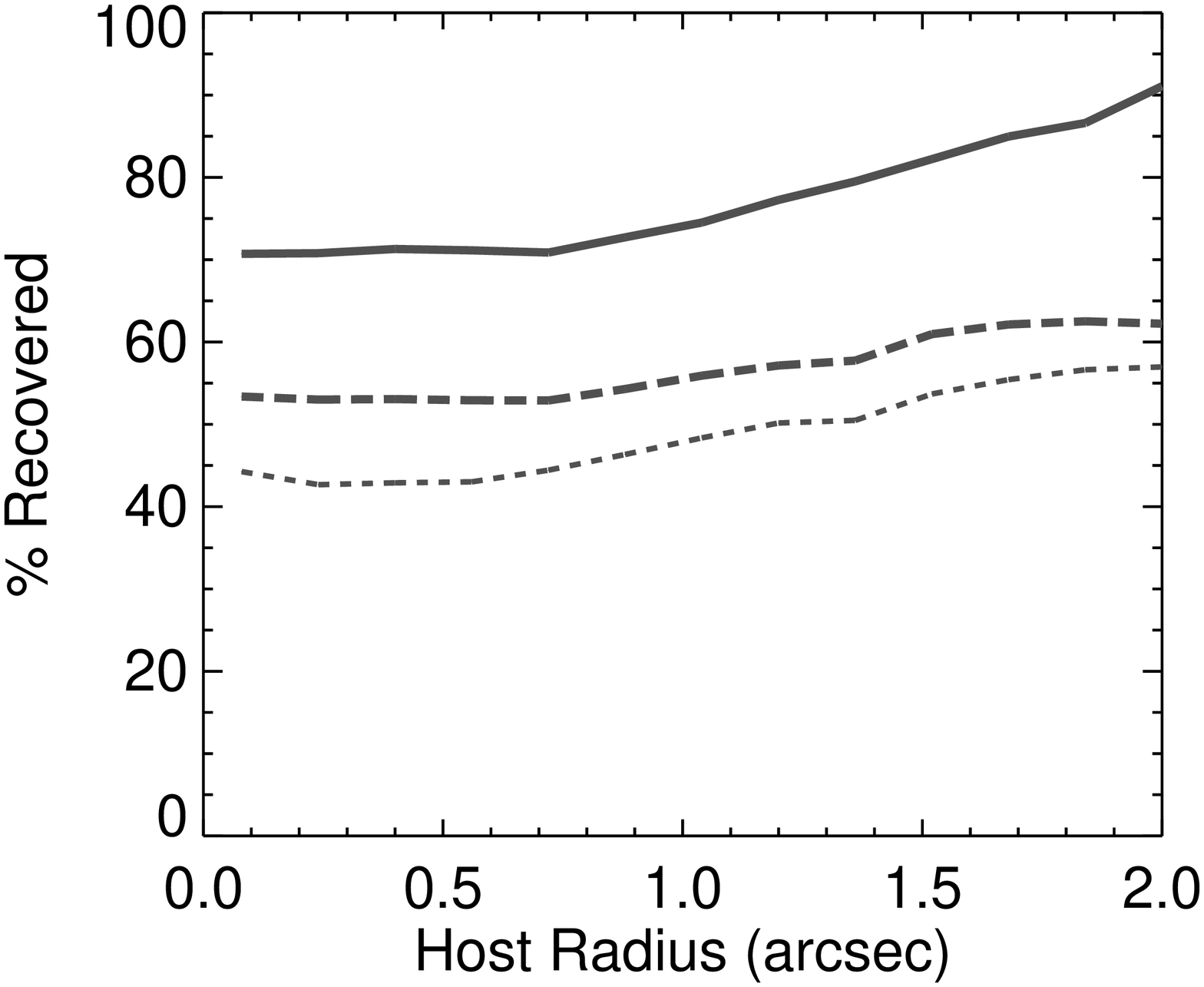}
\includegraphics[width=2.1in]{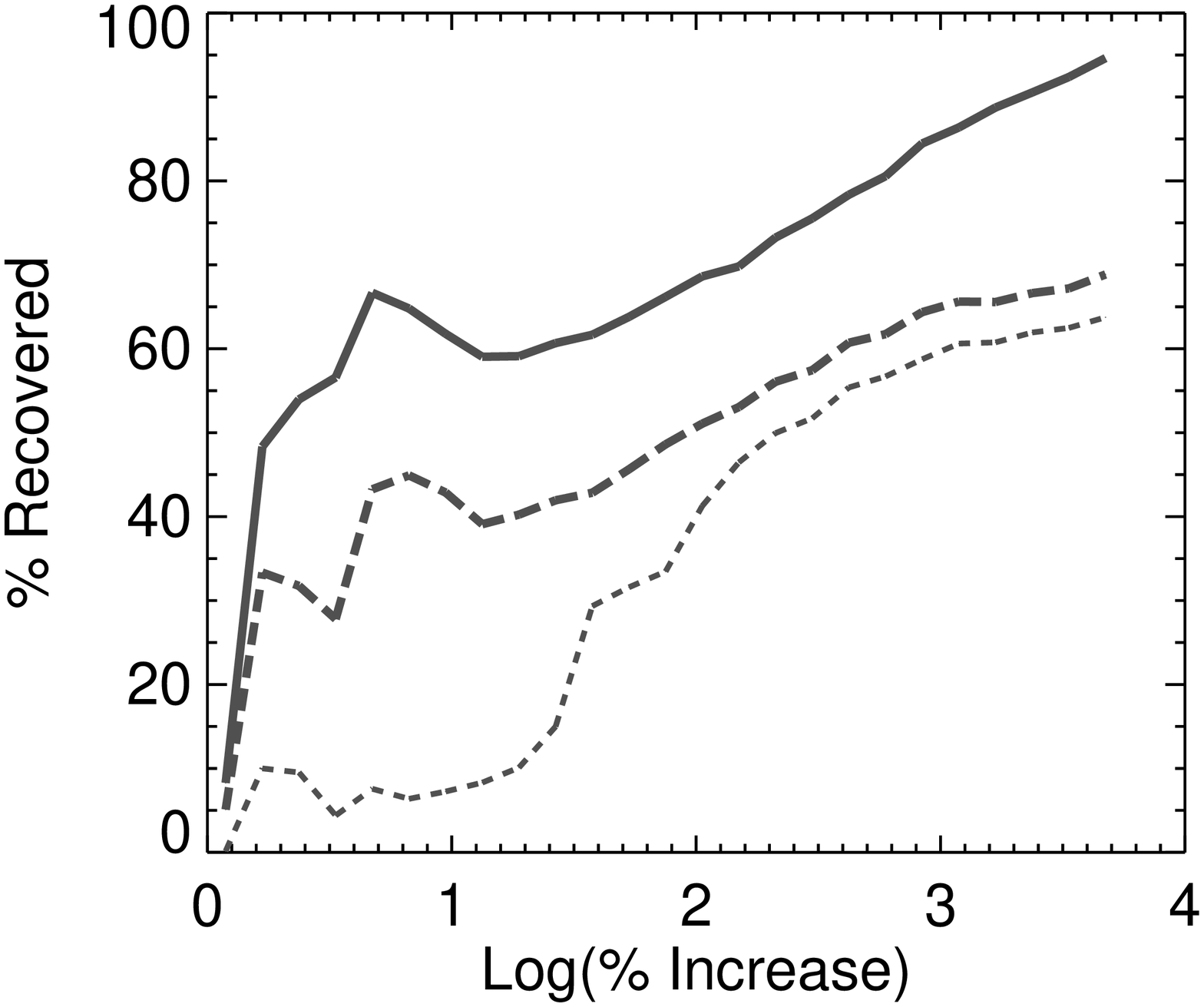}
\caption{SN~Ia recovery as a function of $\Delta$mag (left),
  galactocentric radius in the host (center), and \%inc (right), as in
  Figure~\ref{fig:comp_zsc}.  Note that the detection effects are
  averaged over hosts with a variety of morphological types.  The bar
  at the left of the \%inc histogram represents all objects with
  \%inc$\leq 0$, while the bar at the right includes SNe~Ia with
  Log(\%inc)$\geq 4$.  The large variations in the recovery fraction
  at Log(\%inc)$\la 1$ are due to small sampling statistics in those
  bins.}
\label{fig:comp_dmgal}
\end{figure}

The SN~Ia recovery results are presented as a function of various
parameters in Figures~\ref{fig:comp_mag}--\ref{fig:comp_dmgal}. These
plots include the cumulative data for all realizations performed on
the field-seasons up to the end of D3 observations in June 2007.  In
each of the top panels of the figures, the input (resampled)
distributions are shown as the unfilled histograms and the recovered
populations are shaded in the lightest grey.  The recovery data
naturally include the timing, spacing, and quality of the actual
observations, as we had the opportunity to detect each supernova while
it was visible on any night that SNLS obtained usable images.

The medium-grey histograms show the effects of incorporating a set of
observational constraints on the artificial SNe that are recovered.
These cuts are for the purpose of ensuring adequate light-curve
coverage for fitting, as also described in \citet{per10}.  The
observational requirements are established as:
\begin{itemize}
\item $\ge 1$ early ($-15$d to $+2.5$d) observation in each of $i_M$ and
  $r_M$;
\item $\ge 1$ early ($-15$d to $+5$d) observation in $g_M$;
\item $\ge 1$ observation near peak ($-9$d to $+7$d) in each of $i_M$
  and $r_M$;
\item $\ge 1$ late ($+5$d to $+20$d) observation in either $i_M$ or
  $r_M$.
\end{itemize}
These dates are in effective rest-frame days, corrected for stretch and
redshift:
\begin{equation}
\label{eq:effage}
\mathrm{age_{eff}} = \frac{\mathrm{age_{obs}}}{s(1+z)}.
\end{equation}
Similar cuts have been made in determining reliable efficiencies in
other SN surveys \citep[e.g.,][]{dil08,dil10}.

The lowest, dark filled histograms in each top panel of
Figures~\ref{fig:comp_mag}--\ref{fig:comp_dmgal} all represent the
spectroscopically-viable population.  In this case, the peak
brightness and \%inc limits are applied as described previously in
\S\ref{sec:followup} for follow-up suitability. In addition, the
candidate must have been discovered prior to $+7$ days past peak (rest
frame), and have at least one observation above background both before
and after maximum light.  Due to the limited availability of telescope
time, not all of the suitable candidates would have actually been
observed spectroscopically.

Figure~\ref{fig:comp_mag} shows the recovery fraction as a function of
$i_M$ peak magnitude.  The ability of the SNLS real-time pipeline to
recover SNe~Ia begins to drop off at magnitudes fainter than $i_M\sim
23$~mag, with a 50\% recovery limit at $i_\mathrm{AB}=24.3$~mag.  This
is measured in one individual night of observation; deeper limits can
be achieved by combining multiple epochs of data in offline searches
at the end of a run.  The dip in the recovery fraction seen after the
application of the observational cuts is likely due to the longer
period over which the higher-stretch bright SNe~Ia can be observed.

The redshift distribution for the recovered objects
(Figure~\ref{fig:comp_zsc}-left) is flat out to $z\sim0.5$ and falls
off smoothly to a 50\% incompleteness at $z\sim1$.  The likelihood of
recovering SNe~Ia is very small at $z\ga 1.1$ using this detection
method.  The small rise at low-$z$ in the recovered fraction once the
observational cuts are applied is due to time-dilation effects on the
SN light curves (Eq.~\ref{eq:effage}). Applying the observational
constraints and spectroscopic cuts brings the 50\% limit down to
$z\sim 0.85$, as shown by the dotted line in the lower-left panel of
Figure~\ref{fig:comp_zsc}.

These results appear to be consistent with similar simulations carried
out for the lower-redshift SDSS-II SN Survey \citep{dil08}. The SDSS
search to $z\sim0.4$ with an efficiency of 0.5 (compared with $z\sim1$
for the SNLS, a factor of 2.5 larger), and report essentially 100\%
efficiency to $z\sim0.2$ (compared with $z\sim0.5$ in our
simulations).  The efficiencies in the SDSS are in good agreement with
those from the SNLS, but multiplied by a factor 0.4 on the redshift
axis.

The recovery fraction as a function of stretch
(Figure~\ref{fig:comp_zsc}-center) exhibits an overall rise for
$s>0.8$ which is primarily a color effect: on average, lower-stretch
SNe are redder, fainter, and tend to have smaller recovery
efficiencies.  The dip at $s=0.8$ is an artifact of the break in the
stretch-color relationship of Eq.~\ref{eq:stretch}.  As expected, the
recovery rate shows a clear decrease towards redder colors
(Figure~\ref{fig:comp_zsc}-right).

The efficiency of SN~Ia discovery depends not only on the brightness
of the candidates, but also on the quality of the background
subtraction in the real-time images.  Furthermore, candidates with low
\%inc values (Eq.~\ref{eq:pcinc}) compared with their hosts will
generally not be considered suitable for spectroscopy.
Figure~\ref{fig:comp_dmgal} provides the distributions and recovery
fractions as a function of $\Delta$mag (left), galactocentric radius
(center), and \%inc (right).  The $\Delta$mag recovery shows a modest
decrease towards increasing values due to the reduced chance of
finding intrinsically fainter objects.  The recovery fraction is
constant for positions out to $r\sim1.0\arcsec$ from the host center,
beyond which radius the relative effects of any galaxy subtraction
residuals will begin to diminish and the detection efficiency will
increase. Note that these effects are averaged over hosts of various
morphological types and brightnesses; the PSF convolution and
subtraction of more highly structured galaxies can leave residuals in
the difference images that can conceal detections.  The \%inc plots at
the right of Figure~\ref{fig:comp_dmgal} reveal that the fractional
brightness over the host galaxy background can have a significant
impact on detection recovery, although the effects of candidate
brightness and color are also folded into this effect.  The details of
the host selection process in the artificial SN~Ia simulations
(Eq.~\ref{eq:aplusb}) are found to have little effect on the recovery
results.

\begin{figure}
\includegraphics[width=3in]{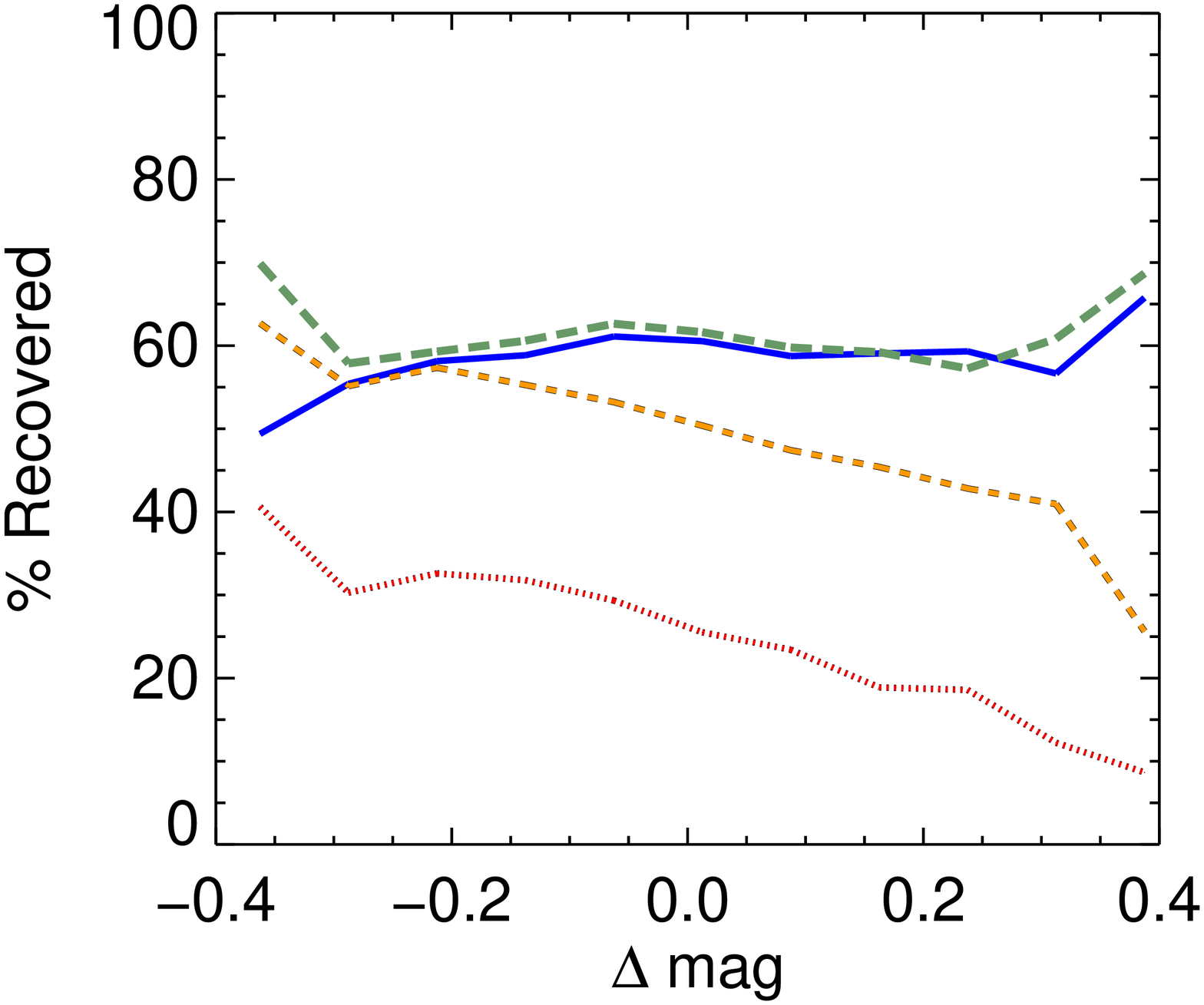}
\includegraphics[width=3in]{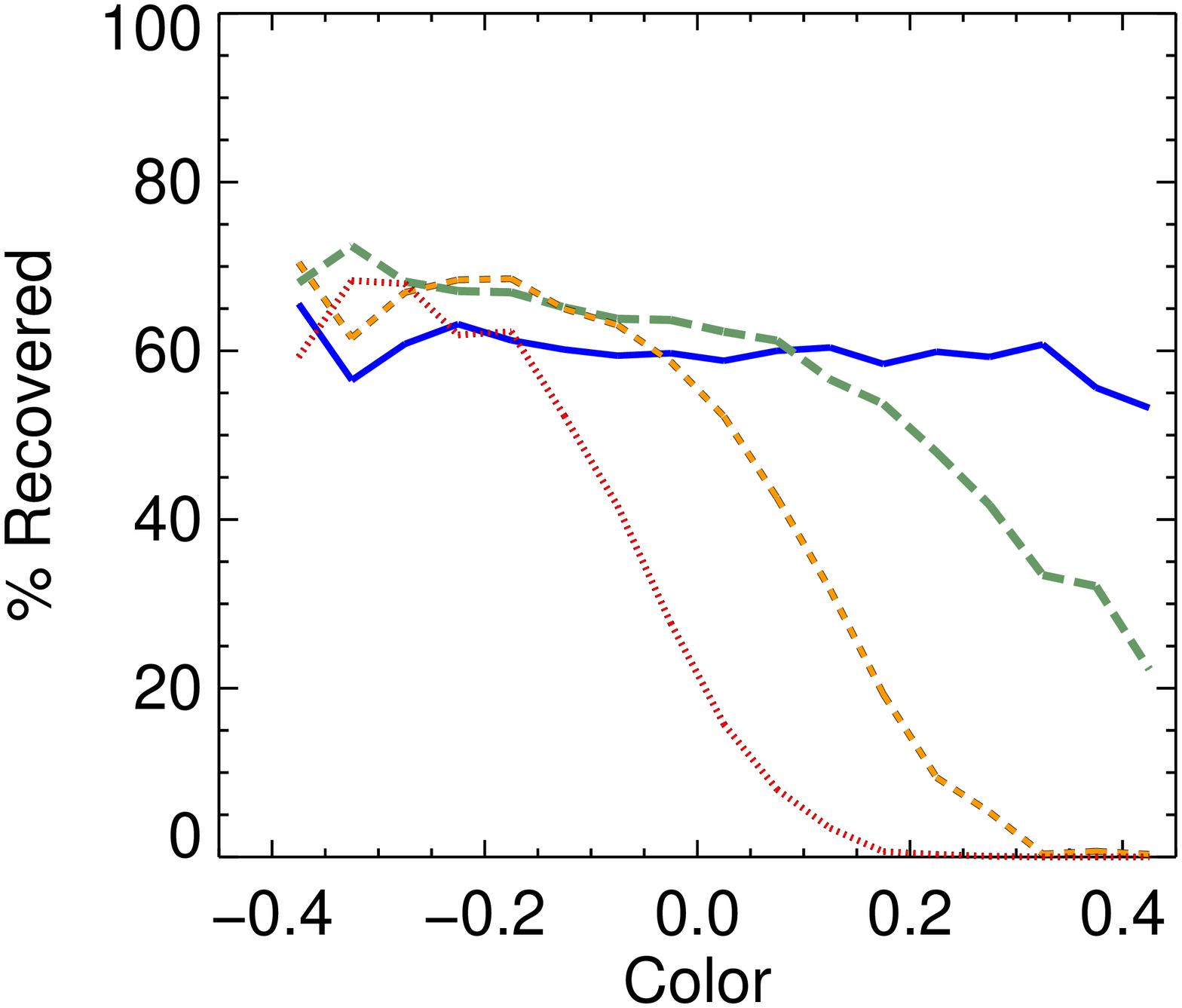}\\
\includegraphics[width=3in]{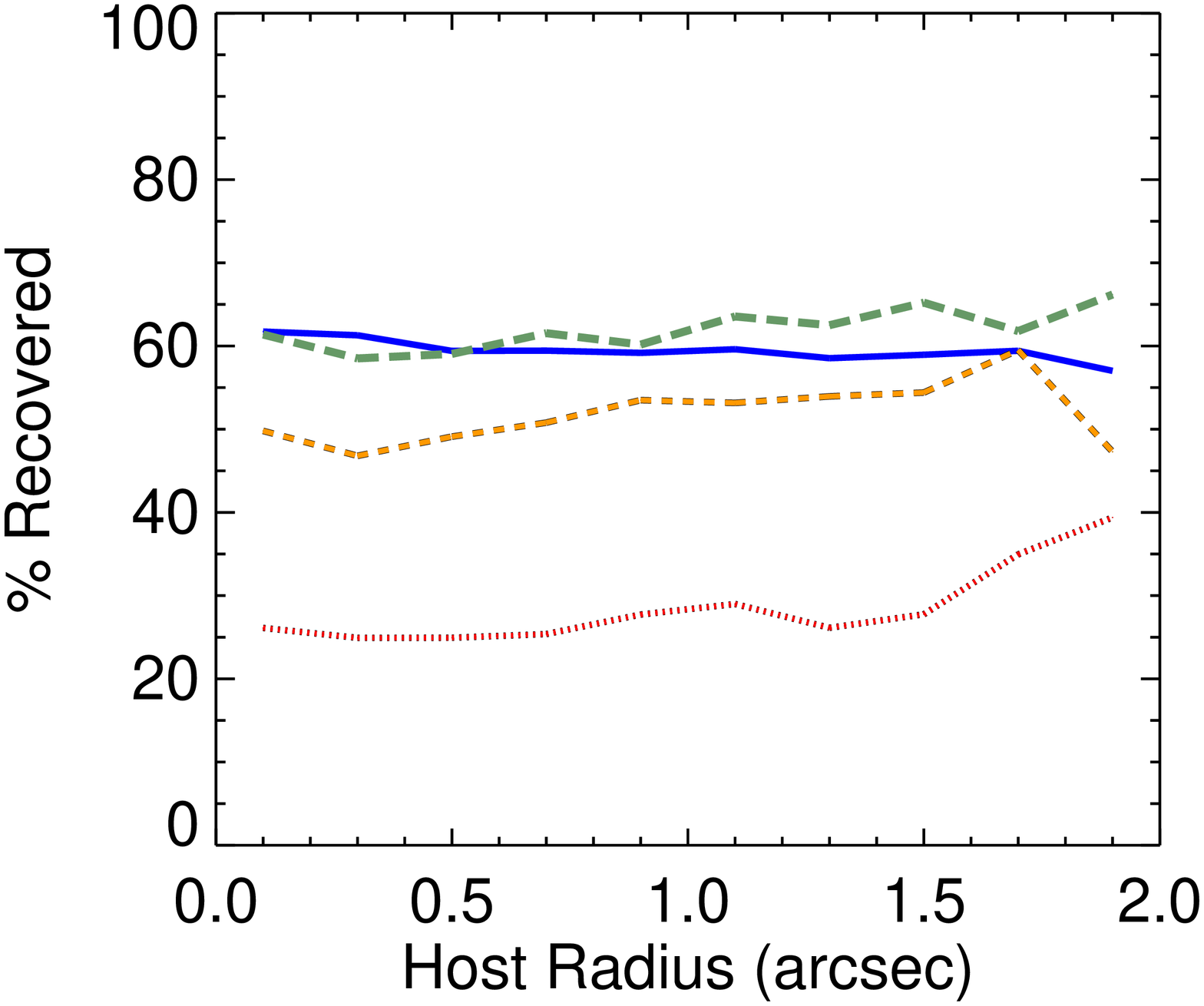}
\includegraphics[width=3in]{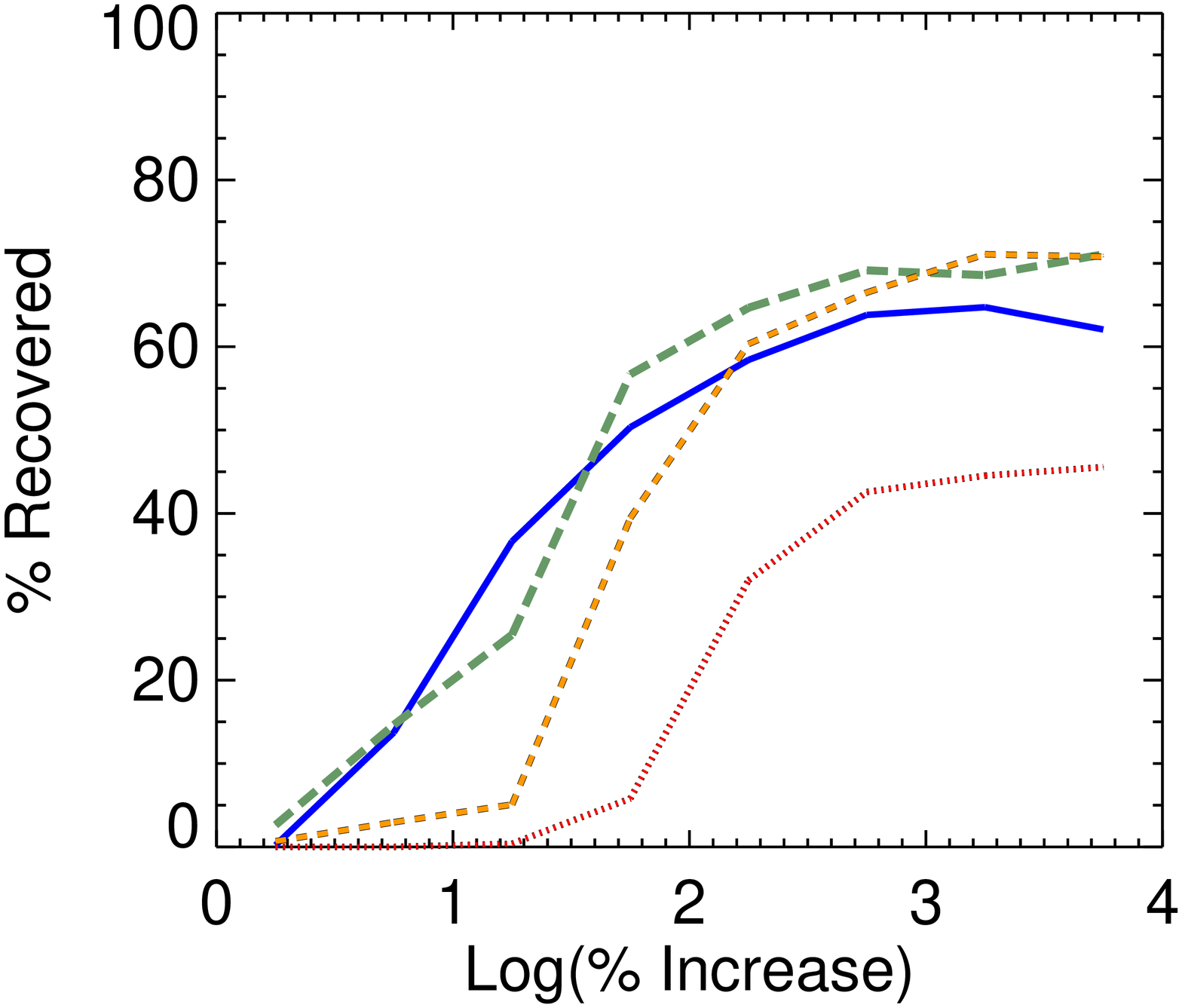}
\caption{SN~Ia recovery fraction for the spectroscopic sample shown
  separately in four redshift bins, given as a function of
  $\Delta$mag, color, galactocentric radius in the host, and \%inc.
  The lines represent the following redshift ranges: $0.10 \leq z <
  0.35$ (solid blue), $0.35 \leq z < 0.60$ (long-dashed green), $0.60
  \leq z < 0.85$ (short-dashed orange), $0.85 \leq z < 1.10$ (dotted
  red).  In each case, the high-redshift range (dotted red line) has
  the lowest recovery rate, due mostly to the peak magnitude and \%inc
  restrictions on the spectroscopic sample.}
\label{fig:comp_zbin}
\end{figure}

Some of the recovery plots in Figures~\ref{fig:comp_zsc} and
\ref{fig:comp_dmgal} can be difficult to interpret in view of the fact
that they incorporate a broad and uniform distribution of redshifts
from $z=0.1-1.2$.  To address this, Figure~\ref{fig:comp_zbin}
provides the spectroscopic recovery fractions in four redshift bins as
a function of $\Delta$mag, $c$, host radius, and \%inc.  In each case,
the incompleteness in the high-redshift bins is the most severe; this
is due primarily to the brightness and \%inc limits imposed on the
spectroscopic sample.

The human review process is another element that must be considered in
the detection recovery estimates; this is complicated by the fact that
fake SNe were not shown to human reviewers during the RTA, as was done
in some surveys \citep[e.g.,][]{dil08}. In the RTA procedure,
candidates were visually confirmed by an operator who could remove
obviously erroneous detections prior to updating the SNLS database.
Since any false detections that made it past the reviewer into the
database could later be culled based on measurements of additional
photometry and subsequent light-curve fits, the only remaining concern
is the omission of good SN~Ia candidates due to human error.

\citet{nei06} performed a Monte Carlo simulation to test the SNLS
human review process using single-epoch images having image quality
(IQ) values of $0\farcs$69 and $1\farcs$09.  Their results indicate
that the Canadian RTA procedure yields a raw recovery percentage
better than $\sim 95\%$ brightward of the incompleteness limit.  Given
multiple chances of candidate discovery using several epochs of SNLS
data, this success rate has been found to approach $\sim 100\%$ over
the duration of a typical, well-observed SNe~Ia. Therefore, any
additional contribution of the visual review process to candidate
recovery is deemed to be insignificant.

\begin{figure}
\plotone{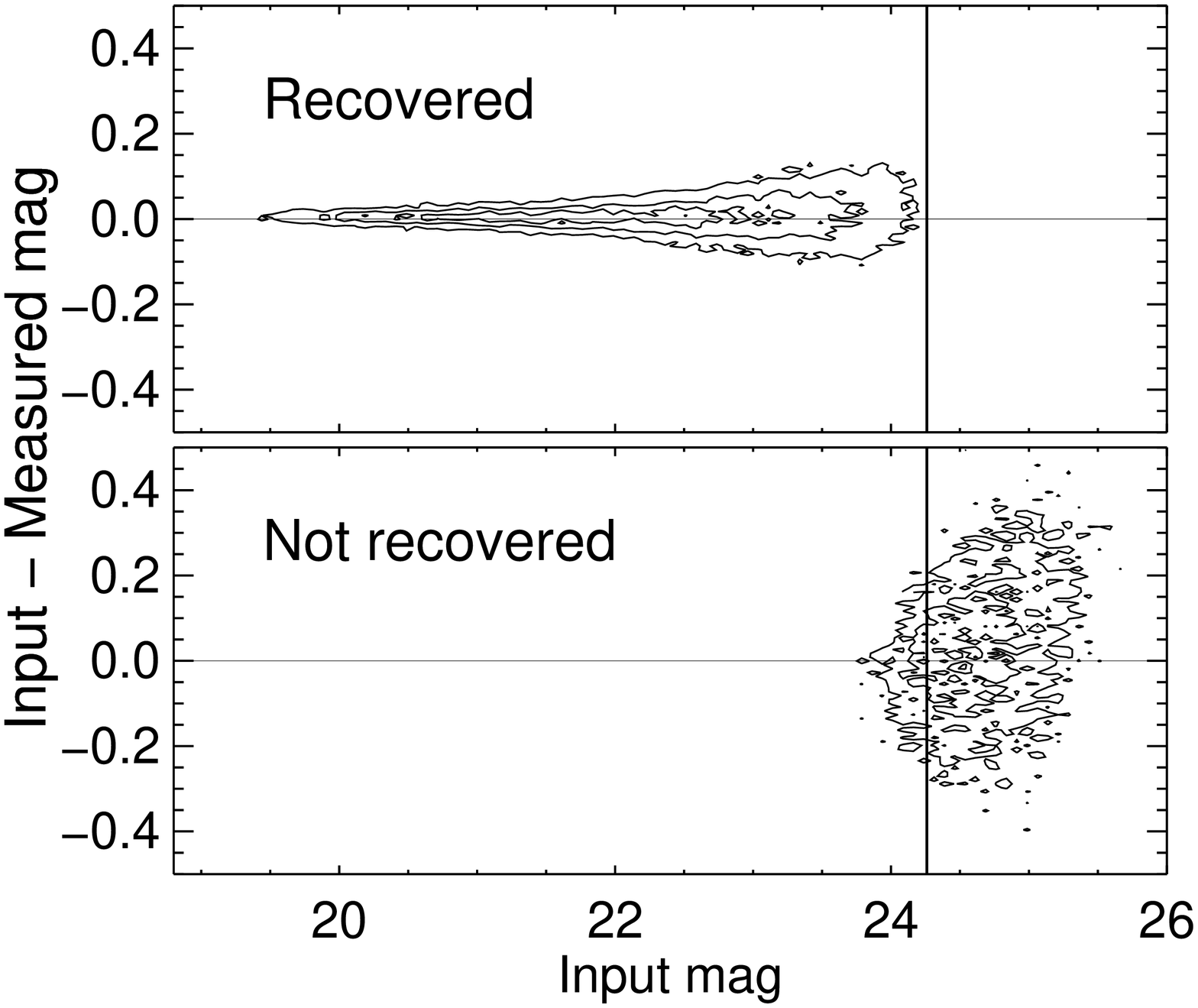}
\caption{Contours showing the difference between input and measured
  magnitude ($i_M$ at peak) versus input magnitude for the recovered
  sample of artificial SN~Ia (top), and those missed by the RTA
  detection pipeline (bottom). The vertical lines show the 50\%
  detection incompleteness limit of $i_M=24.3$~mag (AB).}
\label{fig:comp_phot}
\end{figure}

A comparison of the input and measured photometry for the artificial
SNe~Ia in the simulations is shown in Figure~\ref{fig:comp_phot}.  No
significant systematic biases are found in the measured photometry
using the RTA pipeline. In bright galaxies, \citet{guy10} have
identified a $\sim 2\%$ oversubtraction of host galaxy flux at the
location of the SN when applying a similar method in the final
photometry analysis; this bias is not apparent to within the precision
of the RTA measurements.

\section{Selection bias in the SNLS sample}
\label{sec:malm}

The effects of systematic errors introduced by Malmquist bias and
spectroscopic sample selection (\S\ref{sec:ranking}) are important
considerations for supernova cosmology: they cause a
redshift-dependent offset in the average relative distance moduli from
which the cosmological parameters are determined.  The distance moduli
($\mu_B$) for a Type~Ia SN can be calculated from:
\begin{equation}
\label{eq:cosmo}
\mu_B = m^\ast_B - M + \alpha(s-1) - \beta{c},
\end{equation}
where $m^\ast_B$ is the peak magnitude in the $B$ band, $s$ and $c$
are stretch and color.  The absolute magnitude, $M$, and the
coefficients $\alpha$ and $\beta$ are determined by minimizing the
residuals on the Hubble Diagram \citep{ast06}.

The artificial SNe~Ia from \S\ref{sec:simresults} are used to
calculate the mean offsets in $\Delta$mag, stretch, and color as a
function of increasing redshift.  The mean properties of the observed
spectroscopic sample can be compared with the underlying (unbiased)
population to look for the effects of sampling biases.  The unbiased
population is represented by the input distributions that were
re-weighted to match the underlying distribution as discussed in
\S\ref{sec:sims} (open histograms in the upper panels of
Figures~\ref{fig:comp_mag}--\ref{fig:comp_dmgal}). This, of course,
assumes that there is no significant evolution in the intrinsic
properties of SNe~Ia with redshift.  

The observed spectroscopic sample consists of recovered objects that
were deemed suitable for follow-up.  We also assume that any SNLS
candidates that were observed spectroscopically could be randomly
selected from this latter sample; due to observing time constraints,
not all suitable objects were queued for spectra.  Furthermore, we do
not include a component to address the spectroscopic identification
efficiency.  This would take into account the candidates that were
observed spectroscopically, but which were unidentifiable or
misclassified from their spectra.  Such sample contamination issues
are expected to be relatively minor for bias calculations, yet are
exceedingly difficult to determine precisely \citep[see discussions
  in][]{bal09,con10}.

\begin{deluxetable}{cccc}
\tablewidth{0pt}
\tablecaption{The SNLS sample bias for the magnitude dispersion ($\Delta$mag), SN stretch ($s$) and SN color ($c$).\label{tab:malmbias}}
\tablehead{
  \colhead{$z$} & 
  \colhead{$\delta\langle \Delta\mathrm{mag} \rangle$\tablenotemark{a}} & 
  \colhead{$\delta\langle s \rangle$} & 
  \colhead{$\delta\langle c \rangle$}
}
\startdata
0.33 & $+0.001 \pm 0.001$ & $ 0.007 \pm 0.002$ & $-0.001 \pm 0.001$ \\
0.37 & $+0.000 \pm 0.001$ & $ 0.006 \pm 0.001$ & $-0.002 \pm 0.001$ \\
0.42 & $-0.001 \pm 0.001$ & $ 0.006 \pm 0.001$ & $-0.002 \pm 0.000$ \\
0.47 & $-0.002 \pm 0.002$ & $ 0.007 \pm 0.002$ & $-0.006 \pm 0.002$ \\
0.53 & $-0.003 \pm 0.002$ & $ 0.009 \pm 0.002$ & $-0.009 \pm 0.002$ \\
0.58 & $-0.004 \pm 0.001$ & $ 0.009 \pm 0.003$ & $-0.014 \pm 0.003$ \\
0.62 & $-0.007 \pm 0.001$ & $ 0.011 \pm 0.002$ & $-0.019 \pm 0.002$ \\
0.68 & $-0.009 \pm 0.002$ & $ 0.015 \pm 0.003$ & $-0.027 \pm 0.003$ \\
0.73 & $-0.009 \pm 0.001$ & $ 0.016 \pm 0.004$ & $-0.032 \pm 0.005$ \\
0.77 & $-0.010 \pm 0.002$ & $ 0.017 \pm 0.004$ & $-0.036 \pm 0.003$ \\
0.82 & $-0.013 \pm 0.001$ & $ 0.019 \pm 0.003$ & $-0.043 \pm 0.004$ \\
0.88 & $-0.015 \pm 0.003$ & $ 0.025 \pm 0.003$ & $-0.053 \pm 0.003$ \\
0.92 & $-0.017 \pm 0.003$ & $ 0.030 \pm 0.004$ & $-0.065 \pm 0.003$ \\
0.97 & $-0.024 \pm 0.001$ & $ 0.037 \pm 0.004$ & $-0.073 \pm 0.003$ \\
1.03 & $-0.029 \pm 0.004$ & $ 0.042 \pm 0.002$ & $-0.081 \pm 0.002$ \\
1.07 & $-0.038 \pm 0.003$ & $ 0.053 \pm 0.005$ & $-0.095 \pm 0.002$ \\
1.13 & $-0.054 \pm 0.003$ & $ 0.073 \pm 0.004$ & $-0.110 \pm 0.002$ \\
1.17 & $-0.070 \pm 0.011$ & $ 0.087 \pm 0.010$ & $-0.124 \pm 0.004$ 
\enddata
\tablenotetext{a}{For the parameters $\Delta$mag, $s$ and $c$, the numbers
listed in each redshift bin are the difference, $\delta$, between the
mean of each parameter for the input SN population, and the mean of
the same parameter for the recovered SN sample. The errors quoted are
the errors in the mean.}
\end{deluxetable}

\begin{figure}
\plotone{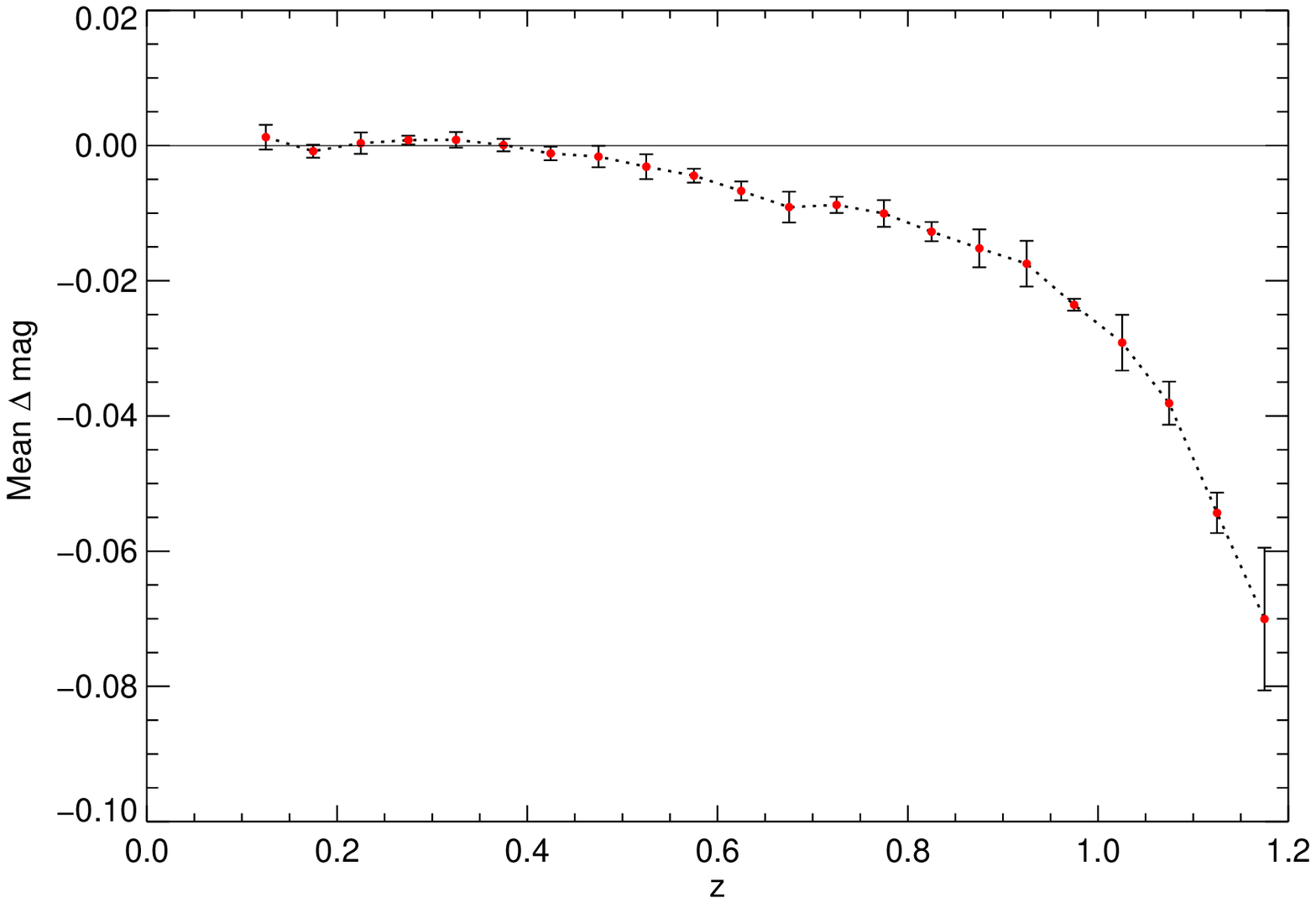}
\caption{Malmquist and spectroscopic selection bias in the SNLS
  sample, as shown by the mean offset in $\Delta$mag as a function of
  redshift.  The errors represent simply the statistical uncertainty
  from the simulations. The small error at $z=0.97$ is a statistical
  fluctuation.}
\label{fig:malm_dmz}
\end{figure}

The offset in the mean $\Delta$mag for the observed sample as a
function of redshift is presented in Figure~\ref{fig:malm_dmz},
showing the combined effects of Malmquist bias and spectroscopic
sampling. The sampling effects start to become significant at $z\ga
0.75$, but are apparent starting at $z \sim 0.6$, shifting the mean
$\Delta$mag in the observed sample towards brighter values.  Column 2
of Table~\ref{tab:malmbias} provides the numerical values of the
average offsets in redshift bins for all fields combined.  The mean
magnitude corrections measured for the individual fields are all
consistent and are listed in Table~\ref{tab:fieldbias}.

\begin{deluxetable}{ccccc}
\tablewidth{0pt}
 \tablecaption{The SNLS sample bias on the magnitude dispersion
($\Delta$mag) in each search field.\label{tab:fieldbias}}
\tablehead{
  \colhead{$z$} & 
  \colhead{$\delta\langle\Delta\mathrm{mag} \rangle_\mathrm{D1}$\tablenotemark{a}} & 
  \colhead{$\delta\langle\Delta\mathrm{mag} \rangle_\mathrm{D2}$} & 
  \colhead{$\delta\langle\Delta\mathrm{mag} \rangle_\mathrm{D3}$} & 
  \colhead{$\delta\langle\Delta\mathrm{mag} \rangle_\mathrm{D4}$}
}
\startdata
0.33 & $+0.002$  & $+0.001$  & $-0.001$  & $+0.001$ \\
0.37 & $-0.001$  & $+0.001$  & $+0.001$  & $+0.000$ \\
0.42 & $-0.001$  & $-0.001$  & $-0.002$  & $-0.000$ \\
0.47 & $-0.000$  & $-0.004$  & $-0.001$  & $-0.001$ \\
0.53 & $-0.001$  & $-0.005$  & $-0.003$  & $-0.004$ \\
0.58 & $-0.004$  & $-0.004$  & $-0.003$  & $-0.006$ \\
0.62 & $-0.006$  & $-0.009$  & $-0.006$  & $-0.006$ \\
0.68 & $-0.008$  & $-0.012$  & $-0.007$  & $-0.010$ \\
0.73 & $-0.009$  & $-0.007$  & $-0.009$  & $-0.010$ \\
0.77 & $-0.007$  & $-0.011$  & $-0.012$  & $-0.010$ \\
0.82 & $-0.013$  & $-0.012$  & $-0.012$  & $-0.015$ \\
0.88 & $-0.012$  & $-0.018$  & $-0.015$  & $-0.016$ \\
0.92 & $-0.017$  & $-0.022$  & $-0.016$  & $-0.015$ \\
0.97 & $-0.024$  & $-0.024$  & $-0.023$  & $-0.023$ \\
1.03 & $-0.023$  & $-0.032$  & $-0.032$  & $-0.030$ \\
1.07 & $-0.038$  & $-0.041$  & $-0.034$  & $-0.040$ \\       
1.13 & $-0.052$  & $-0.053$  & $-0.054$  & $-0.059$ \\
1.17 & $-0.072$  & $-0.084$  & $-0.066$  & $-0.059$ \\
\enddata
\tablenotetext{a}{In each redshift bin, the difference, $\delta$, between the
mean $\Delta$mag for the input SN population and the mean of the same
parameter for the recovered SN sample is given. The errors quoted are
the errors in the mean.}
\end{deluxetable}

\begin{figure}
\plotone{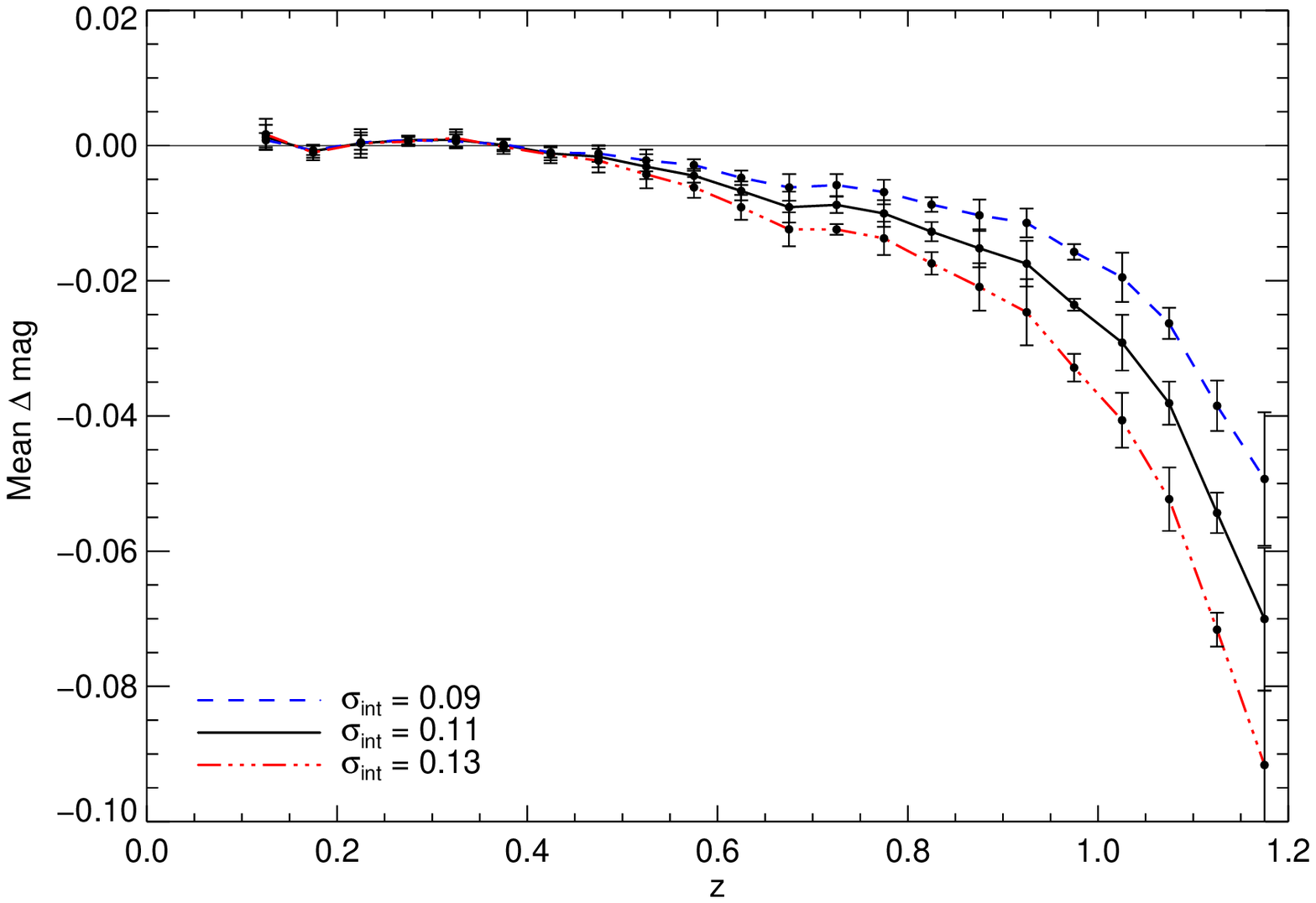}
\caption{The mean offset in $\Delta$mag as a function of redshift, as
  in Figure~\ref{fig:malm_dmz}, shown for three different values of
  $\sigma_\mathrm{int}$.  The actual uncertainty in
  $\sigma_\mathrm{int}$ is around 0.01 mag.}
\label{fig:malm_dmz_intdisp}
\end{figure}

The offset in observed $\Delta$mag with redshift depends strongly on the
intrinsic dispersion of the actual sample.  A comparison of the mean
bias for different assumed values of $\sigma_\mathrm{int}$ in the
artificial population is given in Figure~\ref{fig:malm_dmz_intdisp}.
The lower the dispersion in the $\Delta$mag distribution, the smaller
the effects of selection bias.  This is due to a reduction in the
relative contribution from intrinsically faint input objects near the
detection limit.

The stretch and color trends with redshift were shown in
Figure~\ref{fig:propevolve}.  The increasing offset towards larger
stretch results from preferentially finding the brighter objects with
broader light curves at high $z$ (Column 3 in
Table~\ref{tab:malmbias}).  Similarly, the mean color offsets (Column
4 in Table~\ref{tab:malmbias}) are caused by a bias towards detecting
brighter --- and hence bluer --- objects at $z\ga0.8$.  These stretch
and color offsets are not directly relevant to the cosmology: we
already correct for first-order variations in $s$ and $c$ via the
$\alpha$ and $\beta$ parameters in Eq.\ref{eq:cosmo}, and the
second-order sampling biases that affect the average intrinsic
luminosity of SNe~Ia with redshift are already incorporated into the
$\Delta$mag corrections provided above.

\section{Systematic Errors in the Malmquist Bias}
\label{sec:sys}

There are many parameters in our estimate of the Malmquist bias,
including: the assumed color-color law, the parameters of the
spectroscopic selection model and its form, the values of
$\sigma_\mathrm{int}$, $\alpha$, $\beta$ and $M$, and the measured
underlying stretch and color distributions.  In this section, we
discuss the effects of these terms on the Malmquist bias.

As noted previously, the color-color relation used to warp our SN
template to match the simulated colors does not match our current
understanding of SN behavior.  In order to determine the effects of
this on our results, we re-weighted the simulations to reproduce the
effects of using the SALT2 color law \citep{guy07}.  The effects are
quite small, less than $0.0003$ mag at $z=1$, so we have not applied
this correction to the results.

The uncertainties in the luminosity distance model are 0.01 mag for
$\sigma_\mathrm{int}$, 0.08 for $\alpha$, 0.1 for $\beta$, and 0.025
for $M$\footnote{We have assumed H$_0$ = 70 km/sec/Mpc.  Formally, our
  distance model is only sensitive to some combination of H$_0$ and
  $M$, and it is this quantity that is constrained to 0.025 mag,
  independent of any assumptions about H$_0$.}.  We assign
uncertainties of 0.1 mag to each of the magnitude limits of our
spectroscopic selection model, and 25\% to each of the \%inc
components.  In addition, we consider changing the form of the
spectroscopic model to linearly interpolate between the \%inc limits
as a function of magnitude rather than abruptly changing from one
limit to another.  For the color and stretch models, we vary the
parameters of the distributions until the $\chi^2$ of the fit to the
stretch and color evolution with redshift increases by one.  The
effects of each of these components are given in Table~\ref{tbl:sys}
in terms of the mean Malmquist bias of the 3rd-year SNLS cosmological
sample \citep{guy10} and the bias at $z=1$.  Since the mean and $z=1$
values are 0.0065 and 0.0237 mag, respectively, the total systematic
error is approximately 20\%.

\begin{deluxetable}{lll}
\tablewidth{0pt}
\tablecaption{Systematic uncertainties in the Malmquist bias 
  estimate arising from assumptions in our simulations.\label{tbl:sys}}
\tablehead{
 \colhead{Term} & \colhead{Average systematic} &
 \colhead{Total uncertainty}\\
 \colhead{} & \colhead{uncertainty} & \colhead{at $z=1$}
}
\startdata
 Spec model params      & 0.0006    & 0.0015 \\
 Spec model form        & 0.0002    & 0.0010 \\
 $\sigma_\mathrm{int}$  & 0.0013    & 0.0049 \\
 $\alpha$               & 0.0002    & 0.0005 \\
 $\beta$                & 0.0003    & 0.0008 \\
 $M$                    & 0.0002    & 0.0003 \\
 Stretch Distribution   & $<0.0001$ & 0.0002 \\
 Color Distribution     & 0.0002    & 0.0004 \\
 Color-color law        & $<0.0001$ & 0.0003 \\
\hline
 Total                  & 0.0015    & 0.0053
\enddata
\end{deluxetable}

\section{Summary}
\label{sec:summary}

The SNLS real-time analysis has discovered thousands of supernova
candidates during its first four years of full operation.  The vast
majority of these candidates remain unconfirmed by spectroscopy due to
observing time limitations, although methods for photometric
identification are improving.  Nonetheless, this will remain a major
limitation for the next generation of surveys that could discover an
order of magnitude more SNe, yet may not have access to much more in
the way of spectroscopic time \citep[see][]{how09}.

Supernova surveys all suffer from the effects of selection biases that
must be carefully considered in any analysis \citep[e.g.,][]{kes09}.
Monte Carlo simulations of SNe~Ia in the SNLS real-time pipeline show
that the effects of Malmquist bias and spectroscopic sample selection
begin to affect the distribution of intrinsic magnitudes observed at
$z\ga 0.6$. At $z=1$, the offset in the mean $\Delta$mag reaches
$\sim-0.027$ mag as the survey systematically observes the
intrinsically brighter SNe.  We find approximately 20\% systematic
errors in our estimate of the Malmquist bias based on the uncertainty
in the input model to our simulations.

The SNLS has produced a supernova dataset of unprecedented size and
quality out to $z\sim1.1$.  This large, homogeneous sample of SNe~Ia
is ideal for cosmology \citep{ast06}, the analysis of SN~Ia rates
\citep{nei06,per10,rip10}, and studies of the SN environments
\citep{gra08,car08,sul10}.  Other types of investigations can be made
using the various samples of objects in the SNLS database: e.g.,
cosmology with SNe~IIp \citep{nug06}, core-collapse SN rates
\citep{baz09}, sub-luminous SNe~Ia \citep{gon10}, analyses of unusual
SNe \citep{how06,sch08}, as well as optical variability studies of
AGNs.  In a carefully designed survey, the process of finding
supernovae is quite straightforward.  The real difficulty lies in
determining accurate calibrations and in the management of detailed
systematic errors \citep{ast06,reg09,guy10,con10}. Future supernova
surveys at all redshifts must focus on obtaining more high-quality and
well-characterized data, not simply on detecting more supernovae. That
challenge will demand significantly more in the way of telescope time
and resources.

\section{Acknowledgments}

We wish to express our gratitude to the CFHT staff --- particularly
Pierre Martin, Jean-Charles Cuillandre, Kanoa Withington, Herb
Woodruff, and the Queued Service Observers for their kind efforts on
our behalf.  Our thanks also go out to the support staff at Gemini,
the VLT, and Keck telescopes, as well as to those who provided us with
local technical support: Hugh Zhao, Ross Macduff, Shirley Huang, Mike
Seymour, and Stephenson Yang. Special thanks go to Tom Merrall for his
work in the early days of SNLS, to Santiago Gonz\'alez-Gait\'an for
useful discussions, and to Vanina Ruhlmann-Kleider for the helpful
comments.  We acknowledge the generous support from our funding
agencies: NSERC, CIAR, CNRS, and CEA.  MS acknowledges support from
the Royal Society, and KP is grateful for financial support from NSERC
in the form of a Postdoctoral Fellowship.




\begin{thebibliography}

\bibitem[Astier et al.(2006)]{ast06} Astier, P., et al.\ 2006, \aap,
447, 31

\bibitem[Balland et al.(2009)]{bal09} Balland, C., et al.\ 2009, \aap,
  507, 85

\bibitem[Bazin et al.(2009)]{baz09} Bazin, G., et al.\ 2009, \aap,
  499, 653

\bibitem[Bertin \& Arnouts(1996)]{ber96} Bertin, E., \& Arnouts,
  S.\ 1996, \aaps, 117, 393

\bibitem[Boulade et al.(2003)]{bou03} Boulade, O., et al.\ 2003,
\procspie, 4841, 72

\bibitem[Bronder et al.(2008)]{bro08} Bronder, T.~J., et al.\ 2008,
\aap, 477, 717

\bibitem[Carlberg et al.(2008)]{car08} Carlberg, R.~G., et al.\ 2008,
  \apjl, 682, L25

\bibitem[Conley et al.(2006)]{con06} Conley, A., et al.\ 
2006, \aj, 132, 1707 

\bibitem[Conley et al.(2008)]{con08} Conley, A., et al.\ 2008,
 \apj, 681, 482

\bibitem[Conley et al.(2010)]{con10} Conley, A., et al.\ 2010, in
preparation 

\bibitem[Dahl\'en et al.(2008)]{dah08} Dahl\'en, T., Strolger, L.-G.,
  \& Riess, A.~G.\ 2008, \apj, 681, 462

\bibitem[Dilday et al.(2008)]{dil08} Dilday, B., et al.\ 2008, \apj,
  682, 262

\bibitem[Dilday et al.(2010)]{dil10} Dilday, B., et al.\ 2010, \apj,
  713, 1026

\bibitem[Ellis et al.(2008)]{ell08} Ellis, R.~S., et al.\ 2008, \apj,
  674, 51

\bibitem[Fraser et al.(2008)]{fra08} Fraser, W.~C., et al.\ 2008,
  Icarus, 195, 827

\bibitem[Fukugita et al.(1996)]{fuk96} Fukugita, M., Ichikawa, T.,
  Gunn, J.~E., Doi, M., Shimasaku, K., \& Schneider, D.~P.\ 1996, \aj,
  111, 1748

\bibitem[Garnavich et al.(2004)]{gar04} Garnavich, P.~M., et
  al.\ 2004, \apj, 613, 1120

\bibitem[Gonz\'alez-Gait\'an et al.(2010)]{gon10} Gonz\'alez-Gait\'an,
  S., et al.\ 2010, in preparation 

\bibitem[Graham et al.(2008)]{gra08} Graham, M.~L., et al.\ 2008, \aj,
  135, 1343

\bibitem[Guy et al.(2007)]{guy07} Guy, J. et al.\ 2007, A\&A,
  466, 11 

\bibitem[Guy et al.(2010)]{guy10} Guy, J. et al.\ 2010, \aap, submitted.

\bibitem[Howell et al.(2000)]{how00} Howell, D.~A., Wang, L., \&
  Wheeler, J.~C.\ 2000, \apj, 530, 166

\bibitem[Howell et al.(2005)]{how05} Howell, D.~A., et al.\ 2005, \apj
634, 1190

\bibitem[Howell et al.(2006)]{how06} Howell, D.~A., et al.\ 2006,
  \nat, 443, 308

\bibitem[Howell et al.(2009)]{how09} Howell, D.~A., et al.\ 2009,
  arXiv:0903.1086

\bibitem[Hsiao et al.(2007)]{hsi07} Hsiao, E.~Y., et al.\ 2007, \apj,
  663, 1187

\bibitem[Jha et al.(2007)]{jha07} Jha, S., et al.\ 2007, \apj,
 659, 122

\bibitem[Kessler et al.(2009)]{kes09} Kessler, R., et al.\ 2009, \apjs, 185, 32

\bibitem[Landolt(1992)]{lan92} Landolt, A.~U.\ 1992, \aj, 104, 340

\bibitem[Le Borgne et al.(2004)]{leb04} Le Borgne, D.,
  Rocca-Volmerange, B., Prugniel, P., Lan\c{c}on, A., Fioc, M., \&
  Soubiran, C. 2004, \aap, 425, 881

\bibitem[Magnier \& Cuillandre(2004)]{mag04} Magnier, E.~A., \&
Cuillandre, J.-C. 2004, \pasp, 116, 449

\bibitem[Malmquist(1936)]{mal36} Malmquist, K.~G.\ 1936, Stockholms
  Observatoriums Annaler, 12, 7

\bibitem[Mannucci et al.(2005)]{man05} Mannucci, F., Della Valle, M.,
  Panagia, N., Cappellaro, E., Cresci, G., Maiolino, R., Petrosian,
  A., \& Turatto, M.\ 2005, \aap, 433, 807

\bibitem[Miknaitis et al.(2007)]{mik07} Miknaitis, G., et al.\ 2007, \apj, 666, 674

\bibitem[Neill et al.(2006)]{nei06} Neill, J.~D., et al.\ 2006, \aj,
132, 1126

\bibitem[Nugent et al.(2006)]{nug06} Nugent, P., et al.\ 2006, \apj,
  645, 841

\bibitem[Oke \& Gunn(1983)]{oke83} Oke, J.~B., \& Gunn, J.~E.\ 1983,
  \apj, 266, 713

\bibitem[Perlmutter et al.(1999)]{per99} Perlmutter, S., et al.\ 1999,
  \apj, 517, 565

\bibitem[Perrett et al.(2010)]{per10} Perrett, K., et al.\ 2010, in
  preparation 

\bibitem[Regnault et al.(2009)]{reg09} Regnault, N., et al.\ 2009,
  submitted.

\bibitem[Riess et al.(1998)]{rie98} Riess, A.~G., et al.\ 1998, \aj,
  116, 1009

\bibitem[Riess et al.(2007)]{rie07} Riess, A.~G., et al.\ 2007, \apj,
  659, 98

\bibitem[Ripoche et al.(2010)]{rip10} Ripoche, P., et al.\ 2010, in
  preparation.

\bibitem[Sako et al.(2008)]{sak08} Sako, M., et al.\ 2008, \aj, 135, 348

\bibitem[Scannapieco \& Bildsten(2005)]{sb05} Scannapieco, E., \&
  Bildsten, L.\ 2005, \apjl, 629, L85

\bibitem[Schawinski et al.(2008)]{sch08} Schawinski, K., et al.\ 2008,
  Science, 321, 223

\bibitem[Smith et al.(2002)]{smi02} Smith, J.~A., et al.\ 2002, \aj,
  123, 2121

\bibitem[Sullivan et al.(2006a)]{sul06a} Sullivan, M., et al.\ 2006a,
\aj, 131, 960 

\bibitem[Sullivan et al.(2006b)]{sul06b} Sullivan, M., et al.\ 2006b,
\apj, 648, 868

\bibitem[Sullivan et al.(2010)]{sul10} Sullivan, M., et al.\ 2010,
  \mnras, in press, arXiv:1003.5119

\bibitem[Wood-Vasey et al.(2007)]{wv07} Wood-Vasey, W.~M., et
  al.\ 2007, \apj, 666, 694

\end{thebibliography}
\end{document}